\documentclass[12pt]{article}

\usepackage{amssymb,epsfig}
\newcommand{\qq}{\langle\bar q q\rangle}

\newcommand\ud{\uparrow\downarrow}
\newcommand\du{\downarrow\uparrow}

\newcommand{\beq}[1]{
\begin{equation}\label{#1}}
\newcommand{\eeq}{\end{equation}}
\newcommand{\bea}[1]{
\begin{eqnarray}\label{#1}}
\newcommand{\eea}{\end{eqnarray}}

\newcommand\re[1]{(\ref{#1})}
\newcommand{\derleft}{\stackrel{\gets}{D}}
\newcommand{\derright}{\stackrel{\to}{D}}

\newcommand\lr[1]{{\left({#1}\right)}}

\def\zslash{\rlap/{\mkern-1mu z}}

\renewcommand{\theequation}{\thesection.\arabic{equation}}
\begin{document}

\begin{titlepage}

\begin{flushright}
\begin{tabular}{l}
IPPP/02/40\\
DCPT/02/80\\
 hep-ph/0207307
\end{tabular}
\end{flushright}
\vspace{1.5cm}

\begin{center}
{\LARGE \bf
Photon Distribution Amplitudes in QCD}
\vspace{1cm}

{\sc Patricia\ Ball}${}^1$,
{\sc V.M.\ Braun}${}^2$,
  and
{\sc N.\ Kivel}${}^{2}$
\\[0.5cm]
\vspace*{0.1cm} ${}^1${\it
   IPPP, University of Durham, Durham DH1 3LE, UK
                       } \\[0.2cm]
\vspace*{0.1cm} ${}^2${\it
   Institut f\"ur Theoretische Physik, Universit\"at
   Regensburg, \\ D-93040 Regensburg, Germany
                       } \\[1.0cm]

\vspace{0.6cm}  
\bigskip  
%\centerline{\large \em \today}  
\bigskip  
%\vfill  
\vskip1.2cm
{\bf Abstract:\\[10pt]} \parbox[t]{\textwidth}{
  We develop a consistent technique for the calculation of  
  real photon emission in hard exclusive processes, which is based on the 
  background field formalism and allows a convenient separation 
  of hard electromagnetic and soft hadronic components of the photon.
  The latter ones are related to matrix-elements of light-cone operators 
  in the electromagnetic background field and can be parametrized in terms of
  photon distribution 
  amplitudes. 
  We construct a  
  complete set of photon distribution amplitudes up to and including twist-4,
 for both chirality-conserving and chirality-violating 
  operators. The distribution amplitudes involve several
nonperturbative parameters and, most importantly,
  the magnetic susceptibility of the quark 
  condensate.
  We review and update previous estimates of the susceptibility and 
  also give new estimates of parameters describing
higher-twist amplitudes from QCD sum rules.
}
  \vfill 
{\em Submitted to Nuclear Physics B } 
\end{center}

\end{titlepage}

\thispagestyle{empty}
{\tableofcontents}

\newpage

\addtocounter{page}{-1}
\section{Introduction}  
\setcounter{equation}{0}  

Hard exclusive processes involving photon emission are 
attracting increasing attention. Examples include
transition 
form factors like $\gamma^*\to \pi\gamma$ with one real and 
one virtual photon \cite{massengrab2},  deeply-virtual Compton 
scattering (DVCS) \cite{massengrab3}
and rare radiative 
$B$-decays like $B\to \ell \bar \nu_\ell\gamma$, $B\to \rho\gamma$ and 
$B\to K^*\gamma$ 
\cite{massengrab1}. A specific feature of the QCD 
description of 
such processes is that a real photon contains both a ``pointlike'', 
electromagnetic (EM), and a ``soft'', hadronic, component. This distinction 
is familiar 
from the studies of the photon structure functions in deep inelastic 
scattering (see e.g.\ \cite{photonSF}), but so far has not been studied,
in a systematic way, for exclusive processes. It is the objective of 
this work to develop an efficient formalism for describing either component
of the photon and to give an update of the available information on the 
hadronic one.

We will work in the background (BG) 
field formalism. This technique was pioneered
by Schwinger for QED \cite{Schwinger} and later
rediscovered as a convenient tool in QCD, in particular in connection 
with calculations of the QCD $\beta$-function \cite{Abbott80},
power-corrections to deep-inelastic scattering \cite{SV81} and 
QCD sum rules \cite{DS81,IS84}. An excellent technical review 
can be found in \cite{NSVZ85}. The principal advantage of the BG 
field approach is its explicit gauge-invariance which allows the 
use of different gauges for the soft (classical) and hard (quantum) fields and
also avoids contact terms; instead of the latter, the formalism involves 
new operators containing the EM instead of 
the gluonic field-strength tensor. It also allows an
intuitive physics interpretation which is important by itself. 
%{\em what exactly do you mean by that?}
As an example \cite{IS84}, consider quarks and antiquarks in the QCD vacuum, 
in a constant (electro)magnetic field. In the weak-field limit, the induced  
magnetisation of the vacuum is proportional to the applied field, the 
quark density, the quark electric charge $e_q$ and 
the parameter $\chi$, the so-called magnetic susceptibility of the quark 
condensate:
\beq{chi1}
   \langle 0|\bar q\sigma_{\alpha\beta} q |0\rangle_F =  
    \,e_q\, \chi\, \langle \bar q q\rangle \,F_{\alpha\beta}.
\eeq 
Here $\qq$ is the quark condensate, $F_{\alpha\beta}$ the field-strength 
tensor of the external EM field, and the index $F$ indicates that the VEV is 
taken in the vacuum in the presence of the field $F_{\alpha\beta}$. 
If the magnetic field 
is allowed to vary with a certain frequency, the response becomes sensitive 
to the quark-antiquark separation and gets more complicated. In the limit 
of light-like separations $z^2=0$, which is relevant for hard exclusive 
processes, the magnetic susceptibility has to be substituted by 
a response function $\phi_\gamma$ \cite{BBK88}:
\beq{chi2}
   \langle 0|\bar q(z)\sigma_{\alpha\beta} q(-z)  |0\rangle_F =  
    e_q\, \chi\, \langle \bar q q\rangle
 \int\limits_0^1 \!du\, F_{\alpha\beta}((1-2u)z) \phi_\gamma(u)\,,
\eeq   
where the normalization is chosen such that $\int_0^1 du\, \phi_\gamma(u) =1$.
For a plane-wave configuration
$F_{\alpha\beta}(z) \sim \exp(-iqz),$ $q^2=0$, and in the 
infinite-momentum frame $q_+\to\infty$, the matrix-element on the l.-h.s.\
of \re{chi2} describes the probability amplitude that a real photon 
with momentum $q$ dissociates into a quark-antiquark pair 
at small transverse separation. The function $\phi_\gamma(u)$ can thus 
be identified with a photon distribution amplitude (DA) 
with $u$ and $1-u$ being the momentum fractions carried by the quark and
the antiquark in the photon, respectively, in full 
analogy with DAs of mesons 
\cite{earlyCZ}. To the best of our knowledge, photon 
DAs have been introduced  
and studied for the first time in Refs.~\cite{BBK88}. Some more 
results on higher-twist distributions can be found in \cite{AB95},
while in \cite{PPRWG98} the leading-twist amplitude has been calculated 
using a chiral quark model in the instanton-vacuum.
 
The subject of the present paper is the detailed study of these DAs, 
up to and including twist-4.  
Most of the discussion will refer to asymptotic real photon states, which 
correspond to a plane-wave configuration of the EM BG 
field, whose frequency is, however, not 
assumed to be small. This deviates from the usual procedure of 
expanding in slowly varying (classical) fields and requires special techniques
which have been worked out in Ref.~\cite{BB89}. 
   
The paper is organized as follows: Section~2 is mainly introductory and
summarizes notations and basic features 
of the BG field method. Section~3 contains a detailed discussion 
of the leading-twist photon DA. Section~4
is devoted to photon DAs of twist-3 and 4. Finally, in Sec.~5
we summarize. The paper also contains 
several appendices with results of more technical nature.

\section{The Background Field Method}
\label{sec:2}  
\setcounter{equation}{0}  
\subsection{Gauge-invariance and contact terms}

The basic idea of the BG field method is to modify the quark part 
of the QCD action by including an EM field $B_\mu$ in addition 
to the colour gluon field $A^a_\mu$. In order to preserve  
EM gauge-invariance, the covariant derivative becomes 
\beq{covD}
  D_\mu = \partial_\mu -i g A^a_\mu t^a - i e_q B_\mu\,,
\eeq
where $e_q = e Q_q = \sqrt{4\pi\alpha}\,Q_q$ with 
$Q_{u,c,t}=+2/3$, $Q_{d,s,b}=-1/3$
is the quark electric charge  
and $\alpha =1/137\ldots$ is the fine-structure constant.
The action  reads (for one flavour)
\beq{Sq}
    {\bf S}_q = \int\! d^4x\, \bar q(x)\big[i\!\not\!\!D\, -m_q\big]q(x)
   = {\bf S}_q^{\rm QCD} + e_q \int\! d^4x\, B_\mu(x) j^\mu(x)\,,
\eeq  
$j_\mu = \bar q \gamma_\mu q$ is the vector current. The EM 
field $B_\mu$ is treated as purely classical, i.e.\ it does not  
participate in loops, and is weak, i.e.\ we only consider expressions
linear in $B_\mu$. In Lorentz-gauge, an {\it ingoing}
photon with momentum $q$ corresponds to the  
plane-wave field-configuration
\beq{Bfield}
   B_\mu(x) = e^{(\lambda)}_\mu e^{-iqx}\,.
\eeq 
Here $e^{(\lambda)}_\mu$ is the photon polarization vector with 
$e^{(\lambda)}_\mu q^\mu =0$. 
The corresponding gauge-invariant EM field-strength tensor 
is
\beq{Ffield}
  F_{\mu\nu}(x) = i(e^{(\lambda)}_\mu q_\nu - q_\mu  e^{(\lambda)}_\nu )
                e^{-iqx}\,.
\eeq  
To avoid confusion, we will always write  $F_{\mu\nu}$ for the EM 
and $G_{\mu\nu} = G_{\mu\nu}^a t^a$ for the gluon field-strength tensors, 
respectively.

\begin{figure}[t]
\centerline{\epsfxsize5.0cm\epsffile{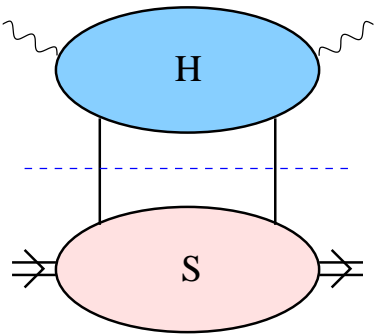}}
\caption[]{
Schematic representation of a hard-soft separation.
 }
\label{fig:hardsoft}
\end{figure}

(Almost) any calculation in QCD relies on the
separation of scales: the contributions of quark and gluon exchanges with 
large momenta are calculated perturbatively and form a hard contribution
(coefficient function), the contributions with small momenta are parametrized
in terms of various distribution functions, alias operator 
matrix-elements. Such a structure is indicated schematically in 
Fig.~\ref{fig:hardsoft}.  
%
%%%%%%%%%%%%%%%%%%     FIGURE 1          %%%%%%%%%%%%%%%%%%%%%%%%%%%%
%%%%%%%%%%%%%%%%%%%%%%%%%%%%%%%%%%%%%%%%%%%%%%%%%%%%%%%%%%%%%%%%%%%%%%
%
A real photon (not shown) can be emitted from both the hard and the 
soft block. The advantage of the BG field method relies on the fact 
that it separates hard and soft photon-emission in an explicitly 
gauge-invariant way.  This is not necessary, but convenient, as
we will see below.% {\em where is that ``below''?}

Technically, the gauge-invariance is seen most directly
if, in the calculation of the hard block, 
one uses the following expression for 
the quark propagator in the BG field\footnote{
For simplicity we present the result for the massless quark
propagator. For massive quarks, see \cite{BBKR94}.}
 \cite{BB89}:
\bea{qprop}
    S(x) %& = & \langle x | \frac{i}{\not\!P} |0\rangle_F\nonumber\\
    &=&
    \frac{i\not\!x}{2\pi^2 x^4}[x,0]_F -
  \frac{i}{16\pi^2x^2} \int\limits_0^1\!\! du\,[x,ux]_F
  \{\bar u \!\not\! x \sigma_{\alpha\beta} + u \sigma_{\alpha\beta}\! \not\!x\}
  e_q F^{\alpha\beta}(ux)[ux,0]_F \nonumber\\
 &&{}+\mbox{\rm terms containing $G_{\alpha\beta}$,}    
\eea  
where $P_\mu = iD_\mu$ is the momentum operator, $\bar u =1-u$ and 
\beq{Pexp}
[x,y]_F ={\rm P}\!\exp\left\{i\!\!\int_0^1\!\! dt\,(x-y)_\mu
  [g A^\mu(tx+ \bar t y)+e_q B^\mu(tx+\bar ty)]\right\}
\eeq
is the path-ordered gauge-link. Here and below the subscript $F$ serves  to 
indicate that the BG field is included, which means in particular  
that both the abelian EM and the nonabelian colour-phases are 
present.%
We will also use the notation $[x,y]$ without a subscript if the
EM field is not included.
 
The terms in $F_{\alpha\beta}$ generate $U(1)$ gauge-invariant 
contributions describing pho\-ton-emis\-sion from small distances, whereas the 
gauge-factors $[.,.]_F$ in the first term on the r.-h.s.\ of \re{qprop}
eventually combine with quark fields that connect the hard and 
the soft block, producing nonlocal gauge-invariant operators. In the simplest 
situation with no external hadrons, one is left with the vacuum expectation
values of such operators in the EM BG field, 
generically
\bea{VEV-F}
   \langle 0| \bar q (x)\Gamma[x,y]_F q(y)|0\rangle_F &=&
    ie_q \int\limits_0^1\! d^4z \,B^\mu(z)\, 
    \langle 0| {\rm T}\{\bar q(x) \Gamma [x,y] q(y) j^\mu(z)\}|0\rangle
\nonumber\\
&& {}+ie_q \int\limits_0^1\! dt\, (x-y)_\mu B^\mu(tx+\bar ty) 
     \langle 0| \bar q(x) \Gamma [x,y] q(y)|0\rangle.
\eea     
Here $\Gamma$ is an arbitrary Dirac-structure. The first term on the r.-h.s.\
of \re{VEV-F} comes from the expansion of the action $e^{i{\bf S}_q}$, 
cf.\ \re{Sq}, whereas the second term originates from the gauge-link
\re{Pexp}. 
The expression \re{VEV-F} is invariant under $U(1)$ gauge-transformations
$B_\mu\to B_\mu +\partial_\mu \phi$. As an example, let us check this 
for plane-waves. Using \re{Bfield}, the r.-h.s.\ becomes 
\bea{check1}
&& i e_q \int\! d^4z \, e_\mu^{(\lambda)} e^{-iqz}  
       \langle 0| {\rm T}\{\bar q(x) \Gamma [x,y] q(y) j^\mu(z)\}|0\rangle
\nonumber\\
   &+& e_q\, e_\mu^{(\lambda)} \frac{(x-y)_\mu}{q(x-y)} 
   \left[e^{-iqy}-e^{-iqx}\right] 
    \langle 0| \bar q(x) \Gamma [x,y] q(y)|0\rangle.
\eea
Note that the expression in the first line is nothing but the amplitude 
of the photon-to-vacuum  transition
\beq{photonstate}
   \langle 0|\bar q(x) \Gamma [x,y] q(y) |\gamma^{(\lambda)}(q) \rangle 
 = i e_q \int\! d^4z \, e_\mu^{(\lambda)} e^{-iqz}  
   \langle 0| {\rm T}\{\bar q(x) \Gamma [x,y] q(y) j^\mu(z)\}|0\rangle.
\eeq 
To check gauge-invariance, replace $e_\mu^{(\lambda)} \to q_\mu$. 
The first line in \re{check1} then becomes 
\beq{check2}
     e_q \int\! d^4z \,e^{-iqz}  \langle 0| \partial_\mu 
      {\rm T}\{\bar q(x) \Gamma [x,y] q(y)  j^\mu(z)\}|0\rangle,
\eeq
which is generally nonzero because of contact terms. To calculate the latter, 
use that inside the QCD functional integral the divergence of the vector 
current can be replaced by a functional derivative over the quark (antiquark)
field:
\beq{byparts}
  \partial_\mu j^\mu(z) \,e^{i{\bf S}_q} = 
  - \bar q(z) \left( \frac{\delta}{\delta \bar q(z)} e^{i{\bf S}_q}\right)
  - \left( \frac{\delta}{\delta q(z)} e^{i{\bf S}_q} \right)q(z)\,. 
\eeq      
Integrating by parts over the quark fields in the functional integral, 
one obtains
\bea{check3}
 & e_q \int\! d^4z \,e^{-iqz} \left\{
  \delta^{(4)}(x-z) \langle 0| \bar q(z) \Gamma [x,y] q(y)|0\rangle 
 - \delta^{(4)}(y-z) \langle 0| \bar q(x) \Gamma [x,y] q(z)|0\rangle\right\}
\nonumber\\
 & = e_q \left[e^{-iqx}-e^{-iqy}\right] 
    \langle 0| \bar q(x) \Gamma [x,y] q(y)|0\rangle\,. 
\eea
This is exactly what is obtained from the 
second term in \re{check1}, but with the opposite sign.
One sees that restoration of $U(1)$ gauge-invariance is achieved   
through elimination of the contact terms.

Calculation of the hard scattering amplitudes 
(hard block in Fig.~\ref{fig:hardsoft}) can usually be done for on-shell 
partons. In the operator product expansion language, this means that 
operators which vanish by virtue of the 
QCD equations of motion (EOM) can be omitted. 
The case of real photon emission is special, because the EOM operators 
can generate nonzero contact terms in correlation functions with the 
EM current. The BG field technique provides an 
important simplification, since no contact terms are present whatsoever, 
but instead the EOM are modified by the presence of the 
BG field. As a simple  example, consider the following 
local operator with two covariant derivatives:
\beq{oper1}
  {\cal O}_{\mu\nu} = \bar q  \gamma_\mu \gamma_\alpha 
    \stackrel{\rightarrow}{D}_{\alpha}\stackrel{\rightarrow}{D}_{\nu} q \,.
\eeq        
For hadronic matrix-elements, one can use the identity 
$[D_\alpha, D_\nu] = -i gG_{\alpha\nu}$ to write
\beq{oper2}
 \langle h'|{\cal O}_{\mu\nu}|h\rangle = -i
 \langle h'| \bar q \gamma_\mu\gamma_\alpha gG_{\alpha\nu}q|h\rangle
 + \langle h'| \bar q \gamma_\mu  
 \stackrel{\rightarrow}{D}_{\nu} \,\stackrel{\rightarrow}{\not\!\! D}\, q|h\rangle 
\eeq  
and neglect the last term. For photons, one either has to take into 
account the EOM operator explicitly, or use the BG field formalism, 
in which case 
there is an additional term in the commutator: $[D_\alpha, D_\nu] 
= -i gG_{\alpha\nu} -i eF_{\alpha\nu}$. We get
\beq{oper3}
  \langle 0|{\cal O}_{\mu\nu}|0\rangle_F = 
   -i\langle 0|\bar q \gamma_\mu\gamma_\alpha gG_{\alpha\nu}q |0\rangle_F 
   -i e_qF_{\mu\nu} \langle \bar q q \rangle 
   + \langle 0|\bar q \gamma_\mu  
 \stackrel{\rightarrow}{D}_{\nu}\, \stackrel{\rightarrow}{\not\!\! D}\, q |\rangle_F \,.
\eeq  
The third term on the r.-h.s.\ is again zero by virtue of the EOM, 
$\not\!\!D\, q =0$,
but there is an additional contribution containing the photon field (and 
the quark condensate $\langle \bar q q \rangle$), which illustrates the generic
feature of the BG field formalism, the appearance of new
operators containing the EM field-strength tensor.

\subsection{Light-cone expansion}
After these general remarks, we proceed to the explicit construction of 
the light-cone expansion of quark-antiquark operators in the BG 
field. We assume that the quark and the antiquark are separated by the 
distance $x_\mu$ and take the limit $x^2\to0$ keeping the scalar 
product $(q\cdot x)$, $q^2=0$, fixed. As discussed in detail in 
\cite{Collins,BB91}, 
the light-cone expansion of a generic nonlocal operator produces
two sequences of terms that are singular and analytic on the light-cone, 
respectively:%
\footnote{Cf.\ in particular Eqs.~(3.7)--(3.11) in \cite{BB91}.}
\beq{anl-sin}
\langle 0| \bar q (x)\Gamma[x,y]_F q(y)|0\rangle_F =
\langle 0| \bar q (x)\Gamma[x,y]_F q(y)|0\rangle^{\rm singular}_F +
\langle 0| \bar q (x)\Gamma[x,y]_F q(y)|0\rangle^{\rm analytic}_F\,.   
\eeq
The singular contributions correspond to small distances, i.e.\ 
to the coefficient function in the OPE expansion, and the analytic 
contributions induce the corresponding operator matrix-elements. The 
separation between singular and analytic contrubutions involves, as usual, 
a factorization scale $\mu_F$. In the case at hand, 
the singular contributions are given by 
\beq{sin}
  \langle 0| \bar q (x)\Gamma[x,y]_F q(y)|0\rangle^{\rm singular}_F =
  - \mbox{\rm Tr}\left[\Gamma[x,y]_F S(y-x)\right]\,, 
\eeq 
where the trace is taken over both  spinor and colour indices and 
$S(y-x)$ is the quark propagator in the BG field \re{qprop}.
As for the analytic contributions, they
can be {\it parametrized} in terms 
of nonperturbative photon DAs of increasing twist. 
To  leading order in the QCD coupling and for massless quarks, one obtains
for the chiral-even operators
\bea{OPEeven}
  \langle 0| \bar q (x)\gamma_\mu[x,-x]_F q(-x)|0\rangle_F &=&
 \int\limits_0^1\!\! du\,\,x^\nu e_qF_{\nu\mu}(-\xi x)\left\{
  -\frac{N_c}{4\pi^2 x^2}\, \xi + 
  {\mathbb G}^{(v)}(u) + {\cal O}(x^2)\right\}, 
\nonumber\\
  \langle 0| \bar q (x)\gamma_\mu\gamma_5 [x,-x]_F q(-x)|0\rangle_F &=&
  i\!\!\int\limits_0^1\!\! du\,\,x^\nu e_q\widetilde F_{\nu\mu}(-\xi x)\left\{
  \frac{N_c}{4\pi^2 x^2} + 
  {\mathbb G}^{(a)}(u) + {\cal O}(x^2)\right\}, 
\eea       
where%
\footnote{The definition of the integration variable $u$ is a convention.
It is chosen such that in the case of an
 {\it ingoing} photon $u$ corresponds to 
the momentum fraction carried by the quark.} 
$$ \xi = 2u-1 $$
and
${\mathbb G}^{(v)}(u)$,  ${\mathbb G}^{(a)}(u)$ are 
nonperturbative functions that depend on $x^2$ at most logarithmically and 
will be  determined later. Here and below we use the 
standard notation 
$\widetilde F_{\alpha\beta} = (1/2) \epsilon_{\alpha\beta\mu\nu} F^{\mu\nu}$ 
and the conventions for the $\epsilon$ tensor and $\gamma_5$ matrix as defined
in \cite{Book}. Counting dimensions, one finds that 
the perturbative contribution in \re{OPEeven} is 
twist-1, and the nonperturbative corrections are twist-3. Note that this 
dimensional twist-counting refers to the ``dynamical'' twist of a 
matrix-element, as opposed to the ``geometric'' twist of a (local) operator.
We will discuss this in more detail in Sec.~4.1.

For chiral-odd operators the perturbative (singular) contribution 
vanishes identically for massless quarks. One then has \cite{AB95}
\bea{OPEodd}
\langle 0| \bar q (x)\sigma_{\mu\nu}[x,-x]_F q(-x)|0\rangle_F &=&
 e_q\langle \bar q q  \rangle
\int\limits_{0}^{1} \!du\,  F_{\mu \nu}(-\xi x)
\left\{
\chi \phi_\gamma(u,\mu)+ \frac{x^2}{4} {\mathbb A} (u)
\right\}
\nonumber\\
&&+
\frac12
e_q\langle \bar q q  \rangle 
\int\limits_{0}^{1} \!du\,
x^\rho\left\{
x^\nu F_{\mu\rho}-x_\mu F_{\nu\rho}
\right\}(-\xi x)\,{\mathbb B}(u) + \ldots,
\nonumber\\
\langle 0| \bar q (x)[x,-x]_F q(-x)|0\rangle_F &=& 0\,. 
\eea 
The function $\phi_\gamma(u,\mu)$ is the (nonperturbative) photon 
DA
of leading twist-2, while the terms in ${\mathbb A} (u)$ 
and ${\mathbb B}(u)$ correspond to twist-4 corrections. 
They will be discussed in detail in the following sections. 
Note that the vacuum
expectation value of the scalar operator does not 
contain any contribution linear in the BG field and thus vanishes 
identically.
  
For the strange quark, and also considering possible lattice 
calculations of DAs, it can be interesting to keep 
the lowest order term in the quark mass $m_q$. 
To leading-twist accuracy one then has
\bea{odd-mass}
 \langle 0| \bar q (x)\sigma_{\mu\nu}[x,-x]_F q(-x)|0\rangle_F &=& 
\int\limits_{0}^{1} \!du\,  e_qF_{\mu \nu}(-\xi x)
\Bigg\{-\frac{N_c m_q}{4\pi^2} \Big[\ln [x^2\mu^2 u(1-u)] +\mbox{\rm const}
\Big]
\nonumber\\
&&{}+ 
\chi\, \qq \phi_\gamma(u,\mu)
\Bigg\}+ \ldots
\eea
(cf.\ App.~A in \cite{BBK88}). Note that in this case there is a weaker, 
logarithmic singularity on the light-cone. The calculation of the constant
under the logarithm requires a careful matching with the UV-divergent 
contribution of the quark loop 
to the photon DA \cite{BBK88}.  
  
It is instructive to verify that the perturbative (singular) contributions
to \re{OPEeven} indeed correspond to the perturbative wave-function 
of the real photon in the infinite-momentum frame. To this end
choose $F_{\mu\nu}$ as a plane-wave, \re{Ffield}, and 
the coordinate system such 
that\footnote{We define $a^\pm=(a^0\pm a^3)/\sqrt{2}$ for any four-vector 
$a$.} $q=(q^+,0,0)$, $x = (0,x_-,\vec{\bf r})$, 
$x^2= -{\bf r}^2$, 
where $\vec{r}$ is a two-dimensional vector in the transverse plane. 
For the leading-twist ``plus'' component, assuming physical (transverse)
photon polarization, one obtains
\beq{wf1}
\langle 0|\bar q(x)\gamma_+\frac{1\!\pm\!\gamma_5}{2}[x,-x]_F q(-x)|0\rangle_F 
= -\frac{iN_c e_q }{8\pi^2 {\bf r}^2} q_+ \!\!\int\limits_0^1 \!
\!du\,e^{i\xi q_+x_-}
\left[(e^{(\lambda)}\cdot {\bf r}) (2u-1) 
 \pm i\epsilon_{ik}{\bf r}_i e^{(\lambda)}_k \right],  
\eeq    
where $\epsilon_{ik}= \epsilon_{ik+-}$ is the two-dimensional antisymmetric 
tensor in the transverse plane, $i,k =1,2$, which is the expected result.
Indeed, the photon wave-function can be defined as   
\bea{wf2}
\psi^{\lambda}_h(u,{\bf r})&=&
\int \frac{dx_{-}}{\pi}e^{-ix_-q_+(2 u-1)}\left.
\langle 0|\bar q(x)\gamma_+\frac{1\!+\!h\gamma_5}{2}[x,-x]_F q(-x)|0\rangle_F
\right|_{x_+=0,x_\perp={\bf r}}
\nonumber \\
&=&-\frac{iN_c e_q }{8\pi^2 {\bf r}^2}
\left\{
(e^{(\lambda)}\cdot {\bf r}) (2u-1) + h\, i
\epsilon_{ik}\, {\bf r}_i \,e^{(\lambda)}_k
\right\}
\eea
where $h=\pm 1$ corresponds to the  quark helicity.  
Choosing  the photon polarisation vector as 
$e^{(\pm)}=(0,1,\pm i,0)/\sqrt{2}$, one obtains the 
familiar expressions given in, for instance,
\cite{Wust97,IW99}\footnote{Up to a different normalization.}.

\section{The Leading-Twist Photon Distribution Amplitude}
\label{sec:3}  
\setcounter{equation}{0}  

We proceed with the detailed study of the 
leading-twist DA of the photon. Like for a 
transversely polarized vector meson \cite{CZreport,BB96}, the 
leading-twist contribution corresponds to the chiral-odd Lorentz-structure. 
Following \cite{BBK88} we define the DA
as the vacuum expectation value (VEV) of the nonlocal quark-antiquark 
operator with light-like separations,
\beq{def1:phi}
   \langle 0|\bar q(z)\sigma_{\alpha\beta}[z,-z]_F q(-z)  |0\rangle_F =  
    e_q\, \chi\, \langle \bar q q\rangle
 \int\limits_0^1 \!du\, F_{\alpha\beta}(-\xi z) \phi_\gamma(u)\,,
\eeq      
where $z^2=0$ and we have inserted the path-ordered gauge-factor which was 
implied, but not shown in~\re{chi2}. This definition is equivalent
to 
\bea{def2:phi}
  i \int\! d^4 y \, e^{-iqy} \langle 0| {\rm T}\{j^\mu(y) 
\bar q(z) \sigma_{\alpha\beta}[z,-z]q(-z)\}|0\rangle
&=&  i(q_\beta g_{\mu\alpha} - q_\alpha g_{\mu\beta})
\chi\, \langle \bar q q\rangle 
  \int\limits_0^1 \!du\, e^{i\xi qz} \phi_\gamma(u)
\nonumber\\
   &&\hspace*{-2.5cm}{} + \frac{z_\mu}{qz} 
   \left[e^{iqz}-e^{-iqz}\right] 
    \langle 0| \bar q(z) \sigma_{\alpha\beta} [z,-z] q(-z)|0\rangle,
\eea  
cf.\ \re{VEV-F} and \re{check1}. The first term on the r.-h.s.\ of \re{def2:phi}
corresponds to the VEV in the BG field of 
the plane-wave, \re{def1:phi}, \re{Ffield}, and the second term is the 
contact term. In fact, in the above case the contact term vanishes: 
the VEV $\langle 0| \bar q(z) \sigma_{\alpha\beta} [z,-z] q(-z)|0\rangle$ 
does not include any external momentum and from $z_\alpha$ alone it is not 
possible to build an antisymmetric Lorentz-structure (in $(\alpha,\beta)$).

Yet another form of the definition \re{def2:phi} is (cf.\ \re{photonstate})
\beq{def3:phi}
   \langle 0 |\bar q(z) \sigma_{\alpha\beta} [z,-z] q(-z) 
   |\gamma^{(\lambda)}(q)\rangle =       
 i \,e_q\, \chi\, \qq
 \left( q_\beta e^{(\lambda)}_\alpha - q_\alpha e^{(\lambda)}_\beta\right)  
 \int\limits_0^1 \!du\, e^{i\xi qz}\, \phi_\gamma(u)\,.
\eeq
 Here we assume that the physical 
photon polarization is transverse to the $(q,z)$ plane. The three definitions
\re{def1:phi}, \re{def2:phi} and \re{def3:phi} are equivalent and can be 
useful in different contexts. 

As always in quantum field theory, 
the extraction of the asymptotic behaviour $z^2\to 0$
creates divergences that have to be regularized. As a result, both 
the magnetic susceptibility $\chi$ and the DA 
$\phi_\gamma(u)$ are scale-dependent. To  leading-logarithmic accuracy: 
\bea{scale}
    \chi(\mu) &=& L^{(\gamma_0-\gamma_{\bar q q})/b}\chi(\mu_0)\,,
\nonumber\\
 \phi_\gamma(u,\mu)  &=& 6u\bar u 
  \left[1+ \sum_{n=2,4,\ldots}^\infty L^{(\gamma_n-\gamma_0)/b}\phi_n(\mu_0) 
  C^{3/2}_n(u-\bar u)\right],
\eea
where $C^{3/2}_n(x)$ are Gegenbauer-polynomials, 
$b=11/3 N_c-2/3 n_f$, $L=\alpha_s(\mu)/\alpha_s(\mu_0)$, 
$\gamma_{\bar q q} = -3 C_F$ is the anomalous dimension of the quark 
condensate and 
\beq{anom-dim}
  \gamma_n = C_F\left[1+4\sum_{j=2}^{n+1}\frac{1}{j}\right]
\eeq
are the anomalous dimensions of the chiral-odd local operators of 
leading-twist \cite{SVy81}. 

Photon DAs are interesting mainly because of the fact that 
the magnetic susceptibility $\chi$ appears to be rather large. A crude 
estimate of its value can be obtained 
in the vector-dominance approximation \cite{BY83}:
consider the correlation function 
\beq{cor-chi}
  i \int\! d^4 y \, e^{-iqy} \langle 0| {\rm T}\{j^\mu(y) 
\bar q(0) \sigma_{\alpha\beta}q(0)\}|0\rangle
 =  i(q_\beta g_{\mu\alpha} - q_\alpha g_{\mu\beta})
\chi(q^2,\mu)\, \langle \bar q q\rangle\,, 
\eeq  
where in contrast to \re{def2:phi} the separation between 
the quarks is set zero and $q^2$ can be arbitrary, instead of 
$q^2=0$ for a real photon. From the normalization of the photon DA
 it follows that $\chi(\mu) \equiv \chi(q^2=0,\mu)$. 
Assuming that the correlation function is dominated by the contribution 
of the lowest-lying states $\rho_0$ and $\omega$ and using the standard 
definitions of couplings\footnote{The currents with specific isospin are 
given by $[\bar q \Gamma q]^{I=1(0)} = (\bar u \Gamma u \mp \bar d \Gamma d)/
\sqrt{2}$. $m_V$ is the average mass of $\omega$ and $\rho(770)$.} 
in the flavour-SU(2) limit
\bea{f-fT}
     \langle 0|[\bar q \gamma_\mu q]^{I=1(0)} |\rho_0 (\omega)\rangle &=& 
      e^{(\lambda)}_\mu f_V m_V\,,
      \qquad\hspace*{52pt} f_V = 215\,\mbox{\rm MeV}~\cite{PDG2000},  
\nonumber\\
     \langle 0|[\bar q \sigma_{\alpha\beta} q]^{I=1(0)} |\rho_0(\omega) 
     \rangle &=& 
     i (e^{(\lambda)}_\alpha q_\beta- e^{(\lambda)}_\beta q_\alpha)
     f^\perp_V\,, \qquad
    f^\perp_V = (160\pm10)\,\mbox{\rm MeV}~\cite{BB96}\,,
\eea  
one obtains 
\beq{VDM1}
     \qq \chi(q^2) \simeq -\frac{m_V f_V f^\perp_V}{m^2_V-q^2}\,,
\eeq
and therefore 
\beq{VDM2}
     \chi_{\rm VDM} \simeq -\frac{f_V f^\perp_V}{m_V\qq} \simeq
        2.7\,\mbox{GeV}^{-2},
\eeq  
using the value $\qq = -(250\,$MeV$)^3$ (at 1 GeV)%
\footnote{The sign of $\chi$ depends on the sign convention used for the EM 
coupling in \re{covD}.}.
Alternatively, one can assume that the vector-dominance approximation 
\re{VDM1} is valid for large Euclidian $q^2\to -\infty$, in which limit the 
leading contribution to the correlation function \re{cor-chi}
is given by the quark condensate:
\beq{Euclid}
   \chi(q^2) \stackrel{q^2\to-\infty}{=} \frac{2}{-q^2}\,.
\eeq 
Comparing  this expression with \re{VDM1}, one obtains an even 
simpler estimate \cite{BY83}
\beq{locdual}
   \chi_{\rm LD} \simeq \frac{2}{m_V^2} \simeq
        3.3\,\mbox{GeV}^{-2}\,, 
\eeq    
which is rather close to \re{VDM2}.
Here the subscript 
``LD'' stands for the so-called local-duality approximation.

A more sophisticated study including contributions of higher 
resonances to \re{VDM1} on one side, and
contributions of higher-dimensional
operators to \re{locdual} on the other side,
combined with the usual QCD sum rule techniques to match between the two
representations in the most efficient way, was done in \cite{BK84,BKY85}.
The result of this analysis,
\beq{SR-chi-old}
   \chi_{\rm SR}(1\,\mbox{GeV}) \simeq 
       4.4\,\mbox{GeV}^{-2}\,,
\eeq
was used in numerous QCD sum rule calculations of EM 
hadron properties over the last 15 years. In App.~B we give an update 
of the analysis of \cite{BK84,BKY85}, using new data for the resonances, 
the most recent
value of $\alpha_s$ and including $O(\alpha_s)$ radiative corrections to the 
sum rule. We obtain 
\bea{SR-chi-new}
   \chi_{\rm SR}(1\,\mbox{GeV}) &\simeq& 
        (3.15\pm 0.3)\,\mbox{GeV}^{-2}\,,%\qquad \mbox{\rm this paper}\,,
 \nonumber\\
   f_\gamma^\perp(1\,\mbox{GeV}) 
&\equiv& \chi \qq = -(50\pm 15)\,\mbox{\rm MeV}\,, 
\eea  
which is smaller compared to the older estimate, mostly due to the large 
negative radiative correction. 

The contribution of the photon DA
 (twist-2) to a generic hard exclusive process 
is, compared to the 
perturbative contribution (twist-1),
suppressed by one power of the hard scale $Q$  and is of order 
\beq{soft/hard}
 \frac{1}{Q}\, \chi\, \frac{(2\pi)^2 |\qq|}{N_c} \approx
 \frac{650\,\mbox{\rm MeV}}{Q}\,.
\eeq  
This relatively large mass scale is due to the large magnetic 
susceptibility and to the well-known 
fact the the quark condensate contribution to physical observables is enhanced 
by a factor $(2 \pi)^2$ compared to the perturbative contributions 
because of a different structure of the phase-space integrals, cf.\ 
Eqs.~\re{OPEeven} and \re{OPEodd}.
  
The available information on the shape of the photon DA 
is more controversial. At large virtualities, the shape is uniquely 
determined by the renormalization properties of the relevant operators,
cf.\ \re{scale}. As the anomalous dimensions in \re{anom-dim}
 rise with $n$, one obtains, for very large scales $\mu$,
the asymptotic photon DA 
\beq{DAasy}
    \phi_\gamma^{\rm asy}(u) = 6 u(1-u),
\eeq    
which is the same as for $\pi, \rho, \ldots$ mesons. In Ref.~\cite{BBK88}
arguments were given suggesting that at smaller virtualities 
of order 1~GeV the photon DA is still close to the 
asymptotic form. The physical picture behind the argumentation in \cite{BBK88}
was that corrections to the asymptotic DA are 
nonperturbative in nature, and are expected to be washed out when one sums
over contributions of different hadron resonances 
$\rho, \rho', \rho^{''},\ldots$. This qualitative conclusion was supported by 
the calculation of the second Gegenbauer-coefficient $\phi_2$ at the scale 
1$\,$GeV using QCD sum rules, yielding a small number. 
Unfortunately, this estimate appears to be very sensitive to the 
choice of input parameters, see App.~B.
In addition, there has been 
increasing evidence in the last years indicating that ``standard'' QCD sum 
rule estimates of corrections to asymptotic DAs
are not reliable. 
We therefore believe that the existing evidence in favour of the 
asymptotic DA of the photon at low scales is
only qualitative.    

\begin{figure}[t]
\centerline{\epsfxsize7.0cm\epsffile{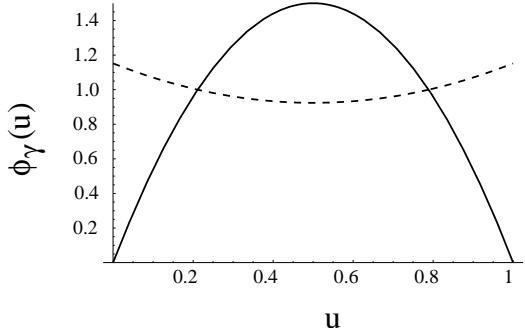}}
\caption[]{ 
The chiral-odd photon DA of leading twist.
The two curves correspond to the asymptotic DA
 \re{DAasy} (solid) and the instanton model at the low
normalization point 0.6~GeV \re{DAIM} (dashed).
% and the instanton model
%including evolution to the normalization point 1~GeV (dash-dotted).
%The evolution is calculated using the one-loop anomalous dimensions 
%\re{anom-dim} and the one-loop running coupling 
%with $\Lambda^{(3)}_{\rm QCD} =200\,$MeV.
%using the value $2\pi/\alpha_s =12$ at the scale 600~MeV 
%adopted in \cite{PPRWG98}.
 }
\label{fig:phigamma}
\end{figure}

Another calculation of the photon 
DA was done in \cite{PPRWG98}, 
using the instanton-model of the QCD vacuum. 
This approach neglects  altogether 
perturbative gluon radiation which generates  
the asymptotic DA, and the results
correspond to a very low normalization point of order of the characteristic 
instanton-size $\rho_I\sim 600\,$MeV. The resulting photon DA
 can to a good accuracy be represented by a simple expression 
\beq{DAIM}
    \phi_\gamma^{\rm IM}(u,\mu = 600\,\mbox{MeV}) = 
\frac{1}{a-1}\left[a - 6 u(1-u)\right]
\eeq    
with $a \simeq 7.58$.\footnote{V.M.B.\ thanks M.\ Polyakov and C.\ Weiss
for providing this 
          parametrization.}
 In the instanton-model 
the shape of the pion and the photon DA turn out 
to be very different. The reason for this 
 is that for the pion there exists  an additional 
(to perturbative radiation) mechanism that suppresses the DA
 at the end-points 
\cite{PPRWG98}. Roughly speaking, this effect is due to 
a nonvanishing size of a constituent quark when it is  probed by  the axial 
current, which is required quite generally by chiral symmetry 
breaking. For the vector current (photon) the constituent quarks appear to
be pointlike, and the suppression is not required. As a result, the 
pion distribution is predicted in this model to be close to the asymptotic 
one already at very low scales, while the photon DA is 
almost flat and does not vanish at the end-points, see Eq.~\re{DAIM} and 
Fig.~\ref{fig:phigamma}.
{}From our point of view this evidence is not conclusive either, since 
the observed end-point behaviour corresponds to contributions of large 
invariant masses, and it is not clear where inclusion of them in the 
instanton-model calculation is legitimate. By its physical meaning, 
the distribution amplitude corresponds to the contributions of low 
momenta and we expect that
an asymptotic-like shape would be recovered if the calculations were made 
with an explicit ultraviolet cutoff of order of the instanton-size.  
%Another problem is that with the currently accepted large value of the 
%strong coupling, effects of the perturbative evolution from the 
%instanton-model scale 600~MeV 
%to the more acceptable 1~GeV are very large, see 
%Fig.~\ref{fig:phigamma}.
We believe that this question will eventually be decided by lattice 
calculations and/or experiments on hard exclusive dijet photoproduction
in which the shape of the photon DA can be measured \cite{photondijets}.  

%
%%%%%%%%%%%%%%%%%%     FIGURE 2          %%%%%%%%%%%%%%%%%%%%%%%%%%%%
%%%%%%%%%%%%%%%%%%%%%%%%%%%%%%%%%%%%%%%%%%%%%%%%%%%%%%%%%%%%%%%%%%%%%%
%         

For the strange quark, one might want to take into account the quark 
mass corrections, in which case there is the complication that the separation 
of  singular and analytic contributions on the light-cone involves
a factorization scale, cf.\ \re{odd-mass}. For completeness, 
we quote here the corresponding result from \cite{BBK88}:
\bea{strange}    
 \lefteqn{  \langle 0|\bar s(x) \sigma_{\alpha\beta}[x,-x] s(-x) 
   |\gamma^{(\lambda)}(q)\rangle =       
   ie_s \left( q_\beta e^{(\lambda)}_\alpha - 
                    q_\alpha e^{(\lambda)}_\beta\right)  \times}\hspace*{0.1cm}
\nonumber\\
 &\times&\int\limits_0^1 \!du\, e^{i\xi qx}
\left\{
     \left(\chi_s \langle \bar s s\rangle - \frac{27 m_s}{g_\phi^2}\right)
     \phi_\gamma(u) %\right. 
%\nonumber\\
%&&\left.{}
-  \frac{3 m_s}{4\pi^2}\left[\ln(-x^2 w_0 u \bar u) -
 \frac{36\pi^2}{g_\phi^2}+2\gamma_E\right]\right\}.
\eea
The quantity $w_0=2$\,GeV$^2$ is the continuum threshold in the classical 
QCD sum rule for the $\phi$-meson \cite{SVZ} and $4\pi^2/g_\phi^2\simeq 11.7$
is the $\phi$-meson coupling (squared) to the EM current. 
By definition, the magnetic susceptibility of the quark condensate does not 
include perturbative contributions of the quark loop $\sim m_s$.

\section{Distribution Amplitudes of Higher Twist}
\label{sec:4}  
\setcounter{equation}{0} 

\subsection{General classification}

The aim of this section is to give a general classification of photon
DAs up to twist-4.  As already stated in Sec.~1,
we define DAs as 
(renormalized) matrix-elements, in the EM BG 
field, of nonlocal gauge-invariant operators 
at strictly light-like separations. 
For brevity, starting from this section we will not show path-ordered
gauge-factors between the fields, they are always implied. We 
distinguish between matrix-elements of operators with an odd number 
of Dirac $\gamma$-matrices, which we term chiral-even, and operators with 
an even number of $\gamma$-matrices, which we refer to as chiral-odd. 

The twist-counting employed in this classification is based on counting 
of powers of the photon momentum in the infinite-momentum frame, cf.\ 
the discussion in \cite{BBKT98}.
To explain this, take a plane-wave configuration
for $F_{\mu\nu}$ and choose the frame of reference such that 
$q_\mu = (q_+,0,{\bf 0}_\perp)$, $z_\mu = (0,z_-,{\bf 0}_\perp)$ 
with $q_+\to \infty$.
The twist-expansion corresponds to counting powers of the large momentum 
$q_+$, taking into account that the photon's transverse polarization vector 
is of order $e^{(\lambda)}_\mu \sim (q_+)^0$ and $z_\mu \sim (q_+)^{-1}$. 
Matrix-elememts of  nonlocal operators typically involve 
several Lorentz-structures which can be separated by suitable projection
operators. In this way, the leading-twist structures  correspond to
the maximum number of ``plus'' components. Each replacement of a ``plus''
projection by a transverse one increases the twist by one unit, and 
each ``minus''  projection adds two units of twist.  
\begin{table}
\addtolength{\arraycolsep}{3pt}
\renewcommand{\arraystretch}{1.4}
$$
\begin{array}{|c|lcc|lc|}
\hline
{\rm Twist} &(\mu) 
& \bar q \gamma_\mu q & 
\bar q \gamma_\mu\gamma_5  q 
&(\mu\nu) & \bar q  \sigma_{\mu\nu} q 
  \\ \hline
2 &  &  & & +\perp &\phi_\gamma 
\\ \hline
3 & \perp &\psi^{(v)}  &\psi^{(a)}  & &\\ 
\hline
4 &  & &  
& \perp - & h_\gamma \\ \hline
\end{array}
$$
\addtolength{\arraycolsep}{-3pt}
\renewcommand{\arraystretch}{1}
\caption[]{
Identification of two-particle DAs with 
projections onto different light-cone components of the 
nonlocal operators. For example, $\perp$ refers to
$\bar q \gamma_\perp\gamma_5 q$.
}\label{tab1}
\end{table}

\begin{table}
\addtolength{\arraycolsep}{2pt}
\renewcommand{\arraystretch}{1.4}
$$
\begin{array}{|c|lcc|lc|lcc|}
\hline
{\rm Twist} &(\mu\nu\alpha) 
& \bar q \widetilde{G}_{\mu\nu}\gamma_\alpha\gamma_5
 q & \bar q G_{\mu\nu}\gamma_\alpha  q 
&(\mu\nu\alpha\beta) & \bar q G_{\mu\nu} \sigma_{\alpha\beta} q 
&(\mu\nu) & \bar q G_{\mu\nu}  q & \bar q \widetilde{G}_{\mu\nu}\gamma_5
 q\\ \hline
3 & +\perp + & {\cal A} & {\cal V} & &&&&\\ \hline
4 &  & & & 
\perp\perp+\!\perp & T_1
& +\perp & S & \widetilde{S}\\
 &  &&&+\perp \perp\perp & T_2 &&&\\ 
& & & & + - +\perp &T_3&&&\\
& & & & +\perp\! +\, - & T_4&&&\\\hline
\end{array}
$$
\addtolength{\arraycolsep}{-3pt}
\renewcommand{\arraystretch}{1}
\caption[]{Same as Tab.~\protect{\ref{tab1}} for 
three-particle DAs. $+\perp\perp$ refers to
$\bar q \widetilde{G}_{+\perp}\gamma_\perp\gamma_5 q$ etc.
}\label{tab2}
\end{table} 

The complete classification of the relevant two- and three-particle 
DAs and their  
relation to specific light-cone projections of the matrix-elements
is given in Tabs.~1 and 2, respectively. As is well known, 
``minus'' components of quark-field operators do not describe independent 
partonic degrees of freedom, but can
 be expressed in terms of multiparton
DAs. The explicit definitions and relations  
between various distributions will be worked out in the next sections.

Note that the classification of photon DAs 
is very similar to the classification of the DAs 
of the transversely polarized $\rho$-meson discussed in  \cite{BBKT98, BB98}.
For the twist-3 distributions considered in the next section 
the analogy is in fact one-to-one. For twist-4 distributions 
the situation is somewhat more complicated because of 
the different structure of the 
QCD EOM in the presence of the BG field. 

\subsection{Twist-3 distribution amplitudes}

The twist-3 DAs of a real photon are related to 
matrix-elements of chiral-even operators.  Since the vacuum expectation 
value of the vector current in the BG EM field vanishes by the EOM, 
$\langle 0|\bar q \gamma_{\mu} q|0\rangle_F =0$, the relevant local operator  
with lowest dimension has to include the gluon field:
\beq{def:f3}
\langle 0|\bar q g\widetilde G_{\mu\nu}\gamma_\alpha\gamma_5 
  q|0\rangle_F  = e_q f_{3\gamma}  D_\alpha F_{\mu \nu},
\eeq 
where $D_\alpha$ is the covariant derivative. The 
nonperturbative constant $f_{3\gamma}$ has dimension GeV$^2$ and 
provides a natural mass-scale for the twist-3 DAs.
There are two two-particle DAs which we define as (cf.\ Tab.~1)
\bea{2F-even}
\langle 0|\bar q (z) \gamma_{\mu} q (-z)|0\rangle_F 
 &= & e_q f_{3\gamma} \int_{0}^{1} \!du\, \bar \psi^{(v)}(u, \mu) F_{z \mu }
(-\xi z)\,, 
\nonumber\\
\langle 0|\bar q(z) \gamma_{\mu} \gamma_{5} q(-z)|0\rangle_F  
&=&e_q f_{3\gamma} \frac{i}2 
\int_{0}^{1} \!du\, \psi^{(a)}(u, \mu)
\widetilde{F}_{z \mu}(-\xi z)\,,
\eea      
where
$$   F_{z \mu } = F_{\rho \mu } z^\rho\,. $$
We will use  similar shorthand notations in what follows. The normalization 
of the DAs is such that $\int_0^1 du \bar\psi^{(v)}(u) = \int_0^1 du 
\psi^{(a)}(u) = 0$. 
Note that the regularization 
(and renormalization) removes the singular $1/z^2$ contributions,
so that the matrix-elements in \re{2F-even} correspond to the analytic parts 
of the amplitudes in the light-cone limit, in the sense of 
the separation in Eq.~\re{anl-sin}.
Comparing \re{2F-even} with \re{OPEeven} we find
\bea{evenLC}
 {\mathbb G}^{(v)}(u) &=&  f_{3\gamma} \bar\psi^{(v)}(u, \mu\sim 1/|x|)\,,
\nonumber\\
 {\mathbb G}^{(a)}(u) &=& \frac12 f_{3\gamma} \psi^{(a)}(u, \mu\sim 1/|x|)\,.
\eea
In addition, there exist two three-particle chiral-even twist-3 distributions: 
\bea{3da}
\langle 0|\bar q(z) g\widetilde G_{\mu\nu}(vz)\gamma_\alpha\gamma_5 
  q(-z)|0\rangle_F  = e_q
  f_{3\gamma} \int {\cal D}\underline{\alpha} \,{\cal A}(\underline{\alpha})
D_\alpha  F_{\mu\nu}(\alpha_v z)\,,
\nonumber\\
\langle 0|\bar q(z) g G_{\mu\nu}(vz)i\gamma_\alpha
  q(-z)|0\rangle_F  = e_q
  f_{3\gamma} \int {\cal D}\underline{\alpha}\, {\cal V}(\underline{\alpha})
D_\alpha F_{\mu\nu}(\alpha_v z)\, ,
\eea
where $\alpha_v=\alpha_{\bar q}-\alpha_{ q}+v\alpha_g$,  
and $\underline{\alpha}$ is the set of the three momentum fractions
$\underline{\alpha}=
\{\alpha_{ q},\alpha_{\bar q},\alpha_g\}$.
 The integration measure is defined as 
\bea{eq:measure}
 \int {\cal D}\underline{\alpha} \equiv \int_0^1 d\alpha_q
  \int_0^1 d\alpha_{\bar q}\int_0^1 d\alpha_g
 \,\delta\left(1-\sum \alpha_i\right).
\eea
Alternatively, the same distributions can be defined as matrix-elements 
between the ingoing photon state and the vacuum, and in this form
they are very similar to the definitions 
adopted in \cite{BBKT98,BB98} for the transversely polarized $\rho$-meson,
apart from a different overall normalization. For two-particle distributions: 
\bea{2even}
\langle 0 |\bar q (z) \gamma_{\mu} q (-z)| \gamma^{(\lambda)}(q)\rangle 
 &= & e_q f_{3\gamma} e^{(\lambda)}_{\perp\mu} 
\int_{0}^{1} \!du\,e^{i\xi qz}\, \psi^{(v)}(u, \mu)\,, 
\nonumber\\
\langle 0|\bar q(z) \gamma_{\mu} \gamma_{5} q(-z)| \gamma^{(\lambda)}(q)
 \rangle 
&=&\frac12e_q f_{3\gamma} \epsilon_{\mu \nu q z}e^{(\lambda)}_{\perp\nu} 
\int_{0}^{1} \!du\, e^{i\xi qz}\,\psi^{(a)}(u, \mu) \,.
\eea  
The functions $\bar\psi^{(v)}(u, \mu) $ and $ \psi^{(v)}(u, \mu)$ 
are related by integration by parts: 
\bea{rel1}
\bar\psi^{(v)}(u, \mu)=2\int_0^u \!d\alpha\, \psi^{(v)}(\alpha, \mu)\,.
\eea 
For the three-particle distributions one has
\bea{eq:3da}
\langle 0
|\bar q(z) g\widetilde G_{\mu\nu}(vz)\gamma_\alpha\gamma_5 
  q(-z)|\gamma^{(\lambda)}(q) \rangle  = e_q f_{3\gamma}  
q_\alpha [q_\nu e^{(\lambda)}_{\perp\mu} 
  - q_\mu e^{(\lambda)}_{\perp\nu}]
\int {\cal D}\underline{\alpha} {\cal A}(\underline{\alpha})
e^{-iqz\alpha_v},
\nonumber\\
\langle 0 |\bar q(z) g G_{\mu\nu}(vz)i\gamma_\alpha
  q(-z)|\gamma^{(\lambda)}(q) \rangle  = e_q f_{3\gamma}  
q_\alpha[q_\nu e^{(\lambda)}_{\perp\mu} 
  - q_\mu e^{(\lambda)}_{\perp\nu}]
\int {\cal D}\underline{\alpha} {\cal V}(\underline{\alpha})
e^{-iqz\alpha_v}.
\eea
%The definitions in \re{2even}, \re{eq:3da} follow the definitions 
%adopted in \cite{BBKT98,BB98} for the transversely polarized $\rho$-mesons
%apart from a different overall normalization and the fact that we choose 
%the photon to be in the final state. 
Like the corresponding $\rho$-meson DAs, the two-particle distributions
$\psi^{(v)}(u)$ and $\psi^{(a)}(u)$ are not independent
and can be related to three-particle distributions using the QCD EOM.
A closer inspection shows that all corrections to the EOM 
in the BG field vanish in the chiral-even case, so that these relations 
remain intact.\footnote{This is in contrast to the chiral-odd ones, to
be discussed in the following subsection, where one VEV is nontrivial,
and induces a nonzero conformal expansion of an exactly vanishing DA.} 
As result, the photon DAs 
can be read off the corresponding expressions for the $\rho$-meson 
\cite{BB98}, 
where one must omit the so-called Wandzura-Wilczek parts that are due to 
the nonzero meson mass. To next-to-leading order in the conformal
expansion, cf.\ Sec.~4.3, we obtain
\bea{evenanswer}
{\cal V} (\underline{\alpha}) &=& 
540\,  \omega^V_\gamma
 (\alpha_{q}-\alpha_{\bar q})\alpha_q \alpha_{\bar q}\alpha_g^2\,,
\nonumber\\
{\cal A} (\underline{\alpha}) &=& 
360\, \alpha_q \alpha_{\bar q} \alpha_g^2 
\left[ 1+ \omega^A_\gamma\,\frac{1}{2}\,(7\alpha_g-3)\right]\,,
\nonumber\\
\psi^{(v)}(u) &=& 5(3\xi^2-1)+ \frac3{64}(
15\omega^V_\gamma-5\omega^A_\gamma )(3-30\xi^2+35\xi^4)\,,
\nonumber\\
\psi^{(a)}(u) &=& (1-\xi^2)(5\xi^2-1)\frac52(1+\frac9{16}\omega^V_\gamma-
\frac3{16}\omega^A_\gamma )\,,
\nonumber\\
\bar\psi^{(v)}(u) &=& -20 u \bar u \xi + \frac{15}{16}\, 
\left(\omega^A_\gamma  
- 3 \omega^V_\gamma\right) u \bar u \xi ( 7 \xi^2-3), 
\eea
where the parameters $\omega^A_\gamma$ and 
$\omega^V_\gamma$ correspond to matrix-elements of local operators
of dimension 6, see \cite{BB98} and App.~B. 

In  vector-dominance approximation one finds
\bea{VDMtwist3}
   f_{3\gamma}^{\rm VDM} &=& - f_\rho^2 \zeta_3 \simeq  -(4 \pm 2)
 \cdot 10^{-3}\,{\rm GeV}^2\,,
\nonumber\\
   \omega^{V, \rm VDM}_\gamma &=& \omega^V_{\rho} \simeq 3.8\pm 1.8\,, 
\nonumber\\
   \omega^{A, \rm VDM}_\gamma &=& \omega^A_{\rho} \simeq -2.1 \pm 1.0\,,
\eea 
where $f_\rho\zeta_3\simeq (20\pm 10)\,$MeV (see App.~B) is the $\rho$-meson 
coupling to the quark-antiquark-gluon current in \re{def:f3}.
Our result for this coupling is by a 
factor 3 larger than an old estimate
in obtained in Ref.~\cite{ZZC}, and quoted in \cite{BBKT98,BB98}. 
We have estimated  corrections to the vector-dominance approximation 
for $f_{3\gamma}$
using QCD sum rules, see App.~B, with the result that these corrections 
are smaller than  the (large) error bars for the $\rho$-meson contribution. 
We, therefore, accept the vector-dominance approximation
for the present. 

The scale-dependence of the twist-3 parameters is given by,
cf.\ Ref.~\cite{BBKT98}, ($C_A = N_c$):
\begin{equation}
 f_{3\gamma} (\mu^2) = L^{\gamma^f/b}f_{3\gamma} (\mu_0^2),
~~~\gamma^f = -\frac{1}{3}C_F +3 C_A,
\end{equation} 
and
\bea{eq:4exam2}
& & \left(
\begin{array}{c}
\omega^{V}_\gamma - \omega^{A}_\gamma\\
\omega^{V}_\gamma + \omega^{A}_\gamma
\end{array}
\right)^{\mu^{2}} =  L^{\Gamma^\omega/b}
\left(
\begin{array}{c}
\omega^{V}_\gamma - \omega^{A}_\gamma\\
\omega^{V}_\gamma + \omega^{A}_\gamma
\end{array}
\right)^{\mu_0^{2}},\nonumber\\
& &\Gamma^\omega = \left(
\begin{array}{cc}
 3C_{F} - {2\over 3}C_{A}\ \ &
\frac{2}{3}C_{F}-\frac{2}{3}C_{A} \\
\frac{5}{3}C_{F} - \frac{4}{3}C_{A}\ \ &
\frac{1}{2}C_{F} + C_{A}
\end{array}
\right).
\eea

\subsection{Twist-4 distribution amplitudes}

\subsubsection{Definitions}

Twist-4 DAs, which are all chiral-odd, are more involved and require 
a detailed discussion.
The general pa\-ra\-me\-tri\-za\-tion
of the nonlocal quark-antiquark operator on the light-cone involves two 
invariant functions:
\bea{2-odd}
\langle 0| \bar q (z)\sigma_{\mu\nu} q(-z)|0\rangle_F &=&
 e_q \chi \qq
\int\limits_{0}^{1} \!du\, \phi_{\gamma}(u,\mu)\, F_{\mu \nu}(-\xi z)
\nonumber\\
&+&{}
\frac12
e_q\qq  
\int\limits_{0}^{1} \!du\, \bar h_{\gamma}(u,\mu)
\left\{
z_\nu F_{\mu z}-z_\mu F_{\nu z}
\right\}(-\xi z)\,,
\nonumber\\
\langle 0| \bar q (z) q(-z)|0\rangle_F &=& 0\,. 
\eea       
Here $\phi_\gamma(u)$ is the leading-twist distribution discussed
in Sec.~3 and $\bar h_{\gamma}(u,\mu)$ is a new DA of twist-4, 
cf.\ Tab.~1. Note that the third possible Lorentz-structure 
$z^\rho\left\{D^\nu F_{\mu\rho}-D_\mu F_{\nu\rho}\right\}$ is not independent 
and can be eliminated using the QED Bianci-identity.

The three-particle DAs are  more numerous and can be defined as:
\bea{def:S}
\langle 0| \bar q(z)g{G}_{\mu\nu}(vz) q(-z) |0 \rangle_F &=&
e_{q}\qq 
\int \!{\cal D}\underline{\alpha}\, {S}(\underline{\alpha})
F_{\mu\nu}(\alpha_v z)\,,
\nonumber\\
\langle 0| \bar q(z)g\tilde{G}_{\mu\nu}(vz)i\gamma_5 q(-z) |0 \rangle_F&=&
e_{q}\qq 
\int\! {\cal D}\underline{\alpha}\, \tilde{S}(\underline{\alpha})
F_{\mu\nu}(\alpha_v z)
\eea
and
\bea{def:TiF}
\lefteqn{
\hspace*{-1cm}\langle 0| \bar q(z)\sigma_{\alpha \beta}g{G}_{\mu\nu}(vz) q(-z) |0 \rangle_F=
      }
\nonumber\\
&=& ie_{q}\qq  \left[g_{\mu \rho}\{g^\perp_{\alpha\nu}D_\beta-
  g^\perp_{\beta\nu}D_\alpha\}- (\mu \leftrightarrow \nu)
               \right]
\int\! {\cal D}\underline{\alpha}\,\bar T_1(\underline{\alpha})
 F_{\rho z}(\alpha_v z)
\nonumber\\
&&{}+
ie_{q}\qq\left[g_{\alpha \rho}\{g^\perp_{\beta\mu}D_\nu- 
  g^\perp_{\beta\nu}D_\mu \}- (\alpha \leftrightarrow \beta)
         \right]
\int\! {\cal D}\underline{\alpha}\,\bar T_2(\underline{\alpha})
 F_{\rho z}(\alpha_v z)
\nonumber\\
&&{}+
ie_{q}\qq \left[ D_\mu z_\nu - D_\nu z_\mu \right]
\int\! {\cal D}\underline{\alpha} \,\bar T_3(\underline{\alpha})
 F_{\alpha\beta}(\alpha_v z)
\nonumber\\
&&{}+ 
ie_{q}\qq \left[ D_\alpha z_\beta - D_\beta z_\alpha \right]
\int\! {\cal D}\underline{\alpha}\,\bar T_4(\underline{\alpha})
 F_{\mu\nu}(\alpha_v z)\,,
\eea
see Tab.~2 for the identification of the relevant light-cone projections.

In addition, we have to consider also quark-antiquark-photon operators:
\beq{phot1}
\langle 0 | \bar q(z)e_q F_{\mu\nu}(vz) q(-z)| 0\rangle_F =
e_q \qq   F_{\mu\nu}(v z)
\eeq
and
\beq{phot2}
\langle 0| \bar q(z)\sigma_{\alpha\beta}\, e_q F_{\mu\nu}(vz)
q(-z)|0\rangle_F = 0.
\eeq
It is convenient to rewrite these expressions introducing auxilary 
three-particle DAs:
\bea{3pform}
\langle 0| \bar q(z)e_q F_{\mu\nu}(vz) q(-z)|0\rangle_F =
e_q \qq 
\int\! {\cal D}\underline{\alpha}\, {S_\gamma}(\underline{\alpha})
F_{\mu\nu}(\alpha_v z),
\eea 
\bea{def:Tg}
\langle 0|\bar q(z)\sigma_{\alpha\beta} e_q F_{\mu\nu}(vz) q(-z)|0\rangle_F =
i e_q\qq  \left[ D_\alpha z_\beta - D_\beta z_\alpha
\right]
\int\! {\cal D}\underline{\alpha}\,\bar T_4^\gamma(\underline{\alpha})
 F_{\mu\nu}(\alpha_v z) +\ldots
\eea
with
\bea{def:SF}
{S_\gamma}(\underline{\alpha})=\delta(\alpha_{\bar q})\delta(\alpha_q), \quad
\bar T_4^\gamma(\underline{\alpha})= 0\,.
\eea
As explained in the next subsection, we need the conformal expansion
of $S_\gamma(\underline{\alpha})$ 
and $\bar T_4^\gamma(\underline{\alpha})$, which turn out to have nonzero 
coefficients.

%As we have seen for twist-2 and twist-3 DAs, 
In some applications it can be 
advantageous to rewrite the definitions \re{2-odd} to \re{def:Tg} 
for the particular case of a photon in the initial state. 
For two-particle DAs we obtain
\bea{2odd-A}
   \langle 0|\bar q(z) \sigma_{\alpha\beta} q(-z) 
   |\gamma^{(\lambda)}(q)\rangle &=&       
 i  \,e_q\, \chi\, \qq
 \left( q_\beta e^{(\lambda)}_\alpha- q_\alpha e^{(\lambda)}_\beta \right)  
 \int\limits_0^1 \!du\, e^{i\xi qz}\, \phi_{\gamma}(u,\mu)
\nonumber\\
&&\hspace*{-1cm}{}+\frac12 i e_q\, \frac{\qq}{qz}
 \left( z_\beta e^{(\lambda)}_\alpha -z_\alpha e^{(\lambda)}_\beta \right)  
 \int\limits_0^1 \!du\, e^{i\xi qz}\, h_{\gamma}(u,\mu)\,.
\eea
It is easy to see that 
\beq{h-hbar}
      \bar h_\gamma (u) = -4\int_0^u d\alpha (u-\alpha) h_\gamma(\alpha,\mu). 
\eeq
C-parity implies the symmetry $\bar h_\gamma(u) = -\bar h_\gamma(1-u)$ 
and $h_\gamma(u) = h_\gamma(1-u)$.

For three-particle distributions the corresponding definitions read
\bea{def:S2}
\langle 0 | \bar q(z)g{G}_{\mu\nu}(vz) q(-z) | \gamma^{(\lambda)}(q) \rangle
&=&
ie_{q}\qq 
[q_\nu e^{(\lambda)}_{\perp\mu}-q_\mu e^{(\lambda)}_{\perp\nu}]
\int {\cal D}\underline{\alpha} {S}(\underline{\alpha})
e^{-i(qz)\alpha_v}\,,
\nonumber\\
\langle 0| 
\bar q(z)g\tilde{G}_{\mu\nu}(vz)i\gamma_5 q(-z) |\gamma^{(\lambda)}(q) \rangle
&=&
ie_{q}\qq [q_\nu e^{(\lambda)}_{\perp\mu} -q_\mu e^{(\lambda)}_{\perp\nu}]
\int {\cal D}\underline{\alpha} \tilde{S}(\underline{\alpha})
e^{-i(qz)\alpha_v},
\eea
and
\bea{def:Ti}
\lefteqn{
\langle 0 | 
\bar q(z)\sigma_{\alpha \beta}g{G}_{\mu\nu}(vz) q(-z) |\gamma^{(\lambda)}(q) \rangle=}
\hspace*{1.6cm}\nonumber\\
&&{} e_{q}\qq 
[ q_\alpha e^{(\lambda)}_{\perp\mu}g^\perp_{\beta\nu}
     -q_\beta e^{(\lambda)}_{\perp\mu}g^\perp_{\alpha\nu}
     -q_\alpha e^{(\lambda)}_{\perp\nu}g^\perp_{\beta\mu}
     +q_\beta e^{(\lambda)}_{\perp\nu}g^\perp_{\alpha\mu} ]
T_1(v,qz)
\nonumber\\
&&{}+
e_{q}\qq 
 [ q_\mu e^{(\lambda)}_{\perp\alpha}g^\perp_{\beta\nu}
     -q_\mu e^{(\lambda)}_{\perp\beta}g^\perp_{\alpha\nu}
     -q_\nu e^{(\lambda)}_{\perp\alpha}g^\perp_{\beta\mu}
     +q_\nu e^{(\lambda)}_{\perp\beta}g^\perp_{\alpha\mu} ]
      T_2(v,qz)
\nonumber \\ 
&&{}+ \frac{e_q\qq}{qz}
[ q_\alpha q_\mu e^{(\lambda)}_{\perp\beta}z_\nu
     -q_\beta q_\mu e^{(\lambda)}_{\perp\alpha}z_\nu
     -q_\alpha q_\nu e^{(\lambda)}_{\perp\beta}z_\mu
     +q_\beta q_\nu e^{(\lambda)}_{\perp\alpha}z_\mu ]
      T_3(v,qz)
\nonumber\\
&&{}+ \frac{e_q\qq}{qz}
    [ q_\alpha q_\mu e^{(\lambda)}_{\perp\nu}z_\beta
     -q_\beta q_\mu e^{(\lambda)}_{\perp\nu}z_\alpha
     -q_\alpha q_\nu e^{(\lambda)}_{\perp\mu}z_\beta
     +q_\beta q_\nu e^{(\lambda)}_{\perp\mu}z_\alpha ]
      T_4(v,qz),\hspace*{1cm}
\eea
where
\bea{Ti}
   T_i (v,qz) =\int {\cal D}\underline{\alpha} 
\,e^{-i(qz)\alpha_v}T_i(\underline{\alpha}).
\eea
The functions $T_i(\underline{\alpha})$ are related to the distributions 
for a generic EM BG field by integration by parts: 
\bea{rel}
\bar T_i(\underline{\alpha})=
-2\int_0^{\alpha_{\bar q}}\!du
\, T_i(\, u,\, \alpha_{\bar q}+\alpha_{q}-u)\,.
\eea
Here we imply $T_i(\underline{\alpha})\equiv T_i(\alpha_{\bar q},\alpha_{q})$.
The corresponding equivalent definitions of $S_\gamma$ and 
$T_4^\gamma$ are analogous to \re{def:S2} and \re{def:Ti}, 
respectively; the relation between $\bar T_4^\gamma$ and $T_4^\gamma$ is
the same as in \re{rel}. 

\subsubsection{Conformal expansion of three-particle twist-4
distribution amplitudes}

In practical calculations one has to use models for the higher-twist DAs
with a few nonperturbative parameters. The major difficulty in 
constructing such models is that the QCD EOM imply the existence of relations 
between different amplitudes and that these relations have to be obeyed 
identically. The standard approach \cite{BBK88,BF90} is to expand the 
amplitudes
in contributions of operators with increasing conformal spin and 
retain only a few leading terms. The reason why this helps is that the
QCD EOM 
are essentially tree-level relations that respect the same symmetries as the 
QCD Lagrangian. Relations between DAs imposed by EOM are therefore 
``horizontal'' in the sense that they relate 
contributions of only the same conformal spin. This implies that the
EOM can be 
solved order by order in the conformal expansion and, most importanly, 
the truncation of the conformal expansion at a certain order is not in 
conflict with the QCD EOM. In addition, as familiar from leading-twist, 
the conformal expansion is consistent with renormalization at LO
level: contributions 
with different conformal spin do not mix with each other to 
leading-logarithmic accuracy. This ensures that higher-spin contributions 
omitted in the model at a certain scale will not reappear at different 
scales through the evolution.     

Technically speaking, the conformal expansion of distribution amplitudes 
is the expansion in orthogonal polynomials that correspond to 
irreducible representations of the conformal group. A suitable basis 
is defined in App.~A. We have already seen examples of conformal 
expansion in Eqs.~(\ref{scale}) and (\ref{evenanswer}).
The expansion of the three-particle DAs of the photon of twist-4
is completely analogous to that of the $\rho$, done in \cite{BB98}.
Without going into details we just quote the results:%
%\footnote{We follow here
%the notations of \cite{BBK88} rather than \cite{BB98}.}
\bea{Tampl}
T_1(\underline{\alpha}) & = & {-}120(3\zeta_2+\zeta_2^+) (\alpha_{\bar q}-\alpha_q)
\alpha_{\bar q}\alpha_q\alpha_g,\nonumber\\
T_2(\underline{\alpha}) & = & 30\alpha_g^2 (\alpha_{\bar q}-\alpha_q)
\left[
 (\kappa-\kappa^+) +\,(\zeta_1-\zeta_1^+)\, (1-2\alpha_g) +
  \zeta_2(3-4\alpha_g)
\right],\nonumber\\
T_3(\underline{\alpha}) & = & -120(3\zeta_2-\zeta_2^+) (\alpha_{\bar q}-\alpha_q)
\alpha_{\bar q}\alpha_q\alpha_g,\nonumber\\
T_4(\underline{\alpha}) & = &
30\alpha_g^2 (\alpha_{\bar q}-\alpha_q) \left[
 (\kappa+\kappa^+) + \,(\zeta_1+\zeta_1^+)\, (1-2\alpha_g) +
  \zeta_2(3-4\alpha_g)
\right],\nonumber\\
{ S}(\underline{\alpha}) & = & 30\alpha_g^2\! \left\{
(\kappa+\kappa^+)\,(1-\alpha_g) +
  (\zeta_1+\zeta_1^+) (1-\alpha_g)(1-2\alpha_g)\right.\nonumber\\
       & & {}\left.+
   \zeta_2[3(\alpha_{\bar q}-\alpha_q)^2-\alpha_g (1-\alpha_g)]\right\} ,\nonumber\\
\widetilde{ S}(\underline{\alpha}) & = & -30\alpha_g^2\! \left\{
 (\kappa-\kappa^+) \,(1-\alpha_g)  +
(\zeta_1-\zeta_1^+) (1-\alpha_g)(1-2\alpha_g)\right.\nonumber\\
& & {}\left.+
   \zeta_2[3(\alpha_{\bar q}-\alpha_q)^2-\alpha_g (1-\alpha_g)]
 \right\}.
\eea
Here terms in $\kappa$ and $\kappa^+$ correspond to the  lowest
conformal spin-3  and the contributions of $\zeta$s are next-to-leading 
with conformal spin-4. For the lowest spin the definitions are:
\bea{def:tS00}
\langle 0| \bar q g{G}_{\mu\nu}  q |0 \rangle_F&=&
e_{q}\qq (\kappa^+ + \kappa)\,
F_{\mu\nu},\nonumber\\
\langle 0| \bar q g\widetilde{G}_{\mu\nu}\gamma_5  q |0 \rangle_F&=&
e_{q} \qq\, (\kappa^+ - \kappa)\,  F_{\mu\nu};
\eea
$\kappa$ and $\kappa^+$ renormalize multiplicatively \cite{BBK88}:
\begin{eqnarray}
\kappa^+(\mu^2) &=& L^{(\gamma^+ - \gamma_{\bar q q})/b}
\kappa^+(\mu_0^2),\quad \gamma_+ = 3 C_A -
\frac{5}{3}\, C_F,\nonumber\\
\kappa(\mu^2) &=& L^{(\gamma^- - \gamma_{\bar q q} /b)}
\kappa(\mu_0^2),\quad \gamma_- = 4 C_A - 3 C_F.
\end{eqnarray}
The $\zeta$-parameters  can be expressed in terms of reduced matrix elements 
$\langle\!\langle Q^{(i)}\rangle\!\rangle$ of local quark-antiquark-gluon
operators with one extra covariant derivative 
and also receive contributions of 
quark-antiquark-photon operators. One finds \cite{BBK88}
\bea{BB:def}
\begin{array}{c}
\displaystyle
\zeta_1=\frac{21}{20}\langle\!\langle
 Q^{(1)}
\rangle\!\rangle,
\mskip10mu
\zeta_2=\frac{7}{22}+\frac7{22}\langle\!\langle
 Q^{(3)}
\rangle\!\rangle-
\frac{21}{220}\langle\!\langle
 Q^{(1)}
\rangle\!\rangle,
\\[4mm]\displaystyle
 \zeta_1^+=7\langle\!\langle
 Q^{(5)}
\rangle\!\rangle,
\mskip10mu
 \zeta_2^+=\frac74\langle\!\langle
 Q^{(3)}
\rangle\!\rangle,
\end{array}
\eea
where $Q^{(1)}$, $Q^{(3)}$, and $Q^{(5)}$ are multiplicatively renormalizable 
operators (in pure QCD, that is not including mixing with 
photon operators) listed in Sec.~4 of Ref.~\cite{BBK88}:
\bea{eq1}
\langle\!\langle Q^{(i)}\rangle\!\rangle (\mu^2) & = &
L^{( \gamma_{Q^{(i)}}-\gamma_{\bar q  q} )/b}\,
\langle\!\langle Q^{(i)}\rangle\!\rangle(\mu_0^2)\nonumber\\
\mbox{with~}\gamma_{Q^{(1)}} &=& \frac{11}2C_A - 3C_F,    \quad
\gamma_{Q^{(3)}} = \frac{13}{3}\, C_F,\quad
 \gamma_{Q^{(5)}} = {5}\,C_A -\frac{8}3 C_F .
\eea
The exact definition of these operators is not important here.
Numerical estimates \cite{BBK88} are collected in Tab.~\ref{tab:para3}.      
%%%%%%%%%%%%%%%%%%%%%%%%%%%%%%%%%%%%%%%%%%%%%%%%%%%%%%%%%%%%%%%%%
\begin{table}
%\renewcommand{\arraystretch}{1.4}
%\addtolength{\arraycolsep}{3pt}
%$$
%\begin{array}{|cccc|}\hline
%f_\rho^T\,[{\rm MeV}] & a_2^\perp & \zeta_3 & \omega_3^T \\ \hline
%160\pm 10 & 0.20\pm 0.10 & 0.032\pm 0.010 & 7.0 \pm 7.0 \\ \hline
%\end{array}
%$$
%\caption[]{Parameters of twist-2 and twist-3
%           chiral-odd distribution amplitudes.
%  Renormalization scale is $\mu = 1\,$GeV.}\label{tab:para2}
\addtolength{\arraycolsep}{3pt}
\renewcommand{\arraystretch}{1.4}
$$
\begin{array}{|c|c|c|c|c|c|}\hline
\kappa  &  \kappa^+ & \zeta_1
 & \zeta_2 & \zeta_1^+ & \zeta_2^+\\\hline 
0.2 & 0 & 0.4 & 0.3 & 0 & 0 \\ \hline
\end{array}
$$
\caption[]{Numerical values of parameters of twist-4
           chiral-odd DAs taken from \cite{BBK88}.
  Renormalization scale is $\mu=1\,$GeV.}\label{tab:para3}
\renewcommand{\arraystretch}{1}
\addtolength{\arraycolsep}{-3pt}
\end{table}

Note that four parameters are defined by three independent matrix elements
so that there is one (exact) relation:
\beq{ferrara}
\zeta_1 + 11 \zeta_2 - 2\zeta_2^+ = \frac72\,,
\eeq
which is a consequence of the theorem by 
Ferrara-Grillo-Parisi-Gatto~\cite{FGPG72}, which states that the divergence 
of a conformal operator vanishes (in a conformally-invariant theory).
In the present case, the theorem implies that the divergence 
of the leading-twist-2 conformal operator with two covariant derivatives
vanishes in free theory and, therefore, in full QCD can be 
expressed in terms of operators including either the gluon or the photon 
field strength. One obtains \cite{BBK88,BB98}
\beq{fer}
\partial_\mu (O^2_{\xi,\eta\alpha\mu} - O^2_{\eta,\xi\alpha\mu}) = 
20(Q^{(4)}_{\alpha,\xi\mu} + Q^{(4F)}_{\alpha,\xi\mu}), 
\eeq
where $O^2_{\perp,+++} = \frac{15}{2}\bar q \sigma_{\perp+}
  \stackrel{\leftrightarrow}{D}_+ \stackrel{\leftrightarrow}{D}_+ q - 
  \frac32\partial_+^2 \bar q \sigma_{\perp+}q$ 
and the explicit expression for the quark-antiquark-gluon operator 
$ Q^{(4)}_{\alpha,\xi\mu}$ is given in Eq.~(4.23) in \cite{BBK88}.
$ Q^{(4F)}_{\alpha,\xi\mu}$
is obtained from $Q^{(4)}_{\alpha,\xi\mu}$ by the 
substitution $gG_{\mu\nu} \to e_qF_{\mu\nu}$.
The vacuum expectation value of the l.-h.s. of 
\re{fer} in the EM field vanishes, whereas 
$\langle\!\langle Q^{(4F)}_{\alpha,\xi\mu}\rangle\!\rangle$ is equal to a constant times
the quark condensate~\cite{BBK88}.
It follows that the reduced matrix element of the complicated quark-antiquark
gluon operator $ Q^{(4)}_{\alpha,\xi\mu}$ in the BG field can be calculated 
exactly in terms of the condensate,
 and this is how  Eq.~\re{ferrara} emerges. This relation is of 
principal importance since it tells that contributions of gluon and photon 
operators cannot be separated in a meaningful way. Hence the use of 
an approximation for gluon contributions necessarily entails the use of the 
corresponding approximation for photon contributions as well.
In particular, the quark-antiquark-photon DAs have to be 
taken into account to the {\it same} accuracy in the conformal 
expansion as the quark-antiquark-gluon ones, notwithstanding that 
the former are known exactly (as given in \re{def:SF}). 

The conformal expansion of quark-antiquark-photon DAs is constructed 
following the approach of  Ref.~\cite{BF90}. To this end one has to
separate different quark-spin projections. We define the 
auxiliary amplitudes
\bea{spinS}
\langle 0 | \bar q(z) \gamma_+ \gamma_- F_{\mu\nu}(vz)
q(-z)|0\rangle & = &
\langle \bar q q\rangle F_{\mu\nu} S^{\ud}(v,pz),
\nonumber\\
\langle 0 | \bar q(z)\gamma_- \gamma_+  F_{\mu\nu}(vz)
q(-z)|0\rangle & = &
\langle \bar q q\rangle F_{\mu\nu} S^{\du}(v,pz).
\eea
Then
\bea{Srel}
{ S_\gamma}(\underline{\alpha}) & = & \frac{1}{2}\, ({
  S}^{\ud}(\underline{\alpha})  +
{ S}^{\du}(\underline{\alpha})),\quad
T_4^\gamma(\underline{\alpha})=\frac{1}{2}\, ({
  S}^{\ud}(\underline{\alpha})  -
{ S}^{\du}(\underline{\alpha}))=0\, .
\eea
The auxiliary amplitudes are expanded in  conformal polynomials as
\bea{sexp}
{ S}^{\ud}(\underline{\alpha}) & = & 60 \alpha_{\bar q}\alpha_g^2 \left
  [ f_{00} + f_{10} \left(\alpha_g-\frac{3}{2}\, \alpha_{\bar q}\right) +
  f_{01} (\alpha_g - 3\alpha_q)\right],\nonumber\\
{ S}^{\du}(\underline{\alpha}) & = & 60 \alpha_q\alpha_g^2 \left
  [ f_{00} + f_{10} \left(\alpha_g-\frac{3}{2}\, \alpha_q\right) +
  f_{01} (\alpha_g - 3\alpha_{\bar q})\right].
\eea
As for the quark-antiquark-gluon DAs, the contributions in $f_{00}$
have conformal spin-3, those in $f_{01}$ and $f_{10}$ conformal
spin-4. 
The constants $f_{ij}$ can be easily obtained by imposing the relations
\bea{prm}
\int {\cal D}\underline{\alpha}\,{S}^{\ud}(\underline{\alpha})=f_{00},\qquad
\int {\cal
D}\underline{\alpha}\,{S}^{\ud}(\underline{\alpha})\alpha_{q,\bar q}=0,
\eea
and using the exact expressions (\ref{def:SF}) for $S_\gamma$ and $T_4^\gamma$.
We find
\bea{fval}
 f_{00}=1\, ,\,  f_{10}=\frac{14}3\, ,\,  f_{01}=\frac73 \, .
\eea
{}With (\ref{Srel}) we then arrive at the following
expressions:
\bea{Sres}
S_\gamma(\underline{\alpha})& =& 30\alpha_g^2\! \left[
 \, [\,(1-\alpha_g) +
   \frac{14}{3}\left\{\!\alpha_g (1-\alpha_g) - \frac{3}{2}\, 
(\alpha_{\bar q}^2 +
  \alpha_q^2) \right\} +\frac73  \left\{ \alpha_g (1-\alpha_g) -
  6\alpha_{\bar q}\alpha_q\right\} \right]
\nonumber\\
& = & 60\alpha_g^2 (\alpha_q+\alpha_{\bar q}) ( 4 - 7 (\alpha_q +
\alpha_{\bar q})),\nonumber\\
T_4^\gamma(\underline{\alpha}) & = & 
30\alpha_g^2 (\alpha_{\bar q}-\alpha_q) \left[
  1 + \frac{7}{3}\, (5\alpha_g-3) +
  \frac{7}{3} \alpha_g\right]\nonumber\\
&=& 60\alpha_g^2 (\alpha_q-\alpha_{\bar q}) 
(4-7(\alpha_q+\alpha_{\bar q})).
\eea
We would like to stress that the conformal expansion of
$T^\gamma_4\equiv0$ is nontrivial:
to any finite order in the expansion one obtains a nonzero result.
This complication arises because $T^\gamma_4$, in contrast to 
the chiral-even twist-3 DAs, does not have simple transformation 
properties under the conformal transformation and corresponds to the 
difference of two amplitudes $S^{\ud}$ and $S^{\du}$ with fixed 
spin projections. The symmetry $S^{\ud} = S^{\du}$ is broken to each 
finite order in the expansion and is only restored in the infinite sum.
Note that if $N$ conformal partial waves are 
taken into account, then the first $N$ moments of $T^\gamma_4$ vanish.
In the next section we will consider an example to illustrate 
why using the truncated expressions in \re{Sres} is mandatory for 
selfconsistency of the OPE. This complication was overlooked in 
\cite{BBK88,AB95} so that the corresponding expressions have to 
be revised.

%Taking into account an infinite
%number of terms, they of course sum up to 0, but to any finite order in
%the expansion one obtains a nonzero result. This is due to the fact
%that $S^{\ud}$ and $S^{\du}$ are nontrivial and is in
%contrast to the chiral-even twist-3 DAs, where {\it all}
%quark-antiquark-photon matrix-elements vanish identically and thus have
%zero conformal expansion.
% This will turn out to be
%important in the next subsection for the calculation of two-particle
%twist-4 DAs.

\subsubsection{Equations of motion and two-particle twist-4
distribution amplitudes}
  
With the three-particle DAs in
hand, we are in a position to calculate the 
two-particle DAs. To
this end, it is convenient to use operator identities for the derivatives 
of the relevant chiral-odd nonlocal operators
and take the vacuum averages in the BG field.
For example, one obtains\footnote{
Here and below we use $\partial_\mu$ to denote the derivative of a 
nonlocal operator with respect
to the total translation, cf.\ \cite{BB89}.} 
\bea{1st}
\partial_\mu \bar  q(x) \sigma_{\mu\nu}  q(-x) & = &
-i\, \frac{\partial}{\partial x_\nu} \,\bar  q(x)
 q(-x) + \int_{-1}^1 dv\, v \bar  q(x)
(gG_{x\nu}+e_q F_{x\nu} )(vx) q(-x)\nonumber\\
& & {} -i\int_{-1}^1 dv\, \bar  q(x) x_\rho (gG_{\rho\mu}+e_q 
F_{\rho\mu} )(vx)
\sigma_{\mu\nu} q(-x),
\nonumber\\
\frac{\partial}{\partial x_\nu} \,\bar  q(x)
\sigma_{\mu\nu}  q(-x) & = & -i\partial_\nu \bar  q(x) q(-x) +
\int_{-1}^1 dv\, \bar  q(x)(gG_{x\nu}+
e_q F_{x\nu} )(vx)
 q(-x)\nonumber\\
& & {} -
i\int_{-1}^1 dv\, v \bar  q(x) (gG_{x\mu}+ e_q F_{x\mu})(vx)
\sigma_{\mu\nu}  q(-x).
\label{2nd}
\eea
These identities can be obtained from the ones derived in
\cite{BBKT98,BB98} by the replacement $gG\to gG + e_q F$. 
Taking matrix-elements 
and comparing with the definitions in (\ref{OPEodd}) and
(\ref{2odd-A}), one obtains the following relations
\bea{1sol}
h_\gamma(u)&=&
-\frac{d}{du}\int_0^u d\alpha_q\int_0^{\bar u} d\alpha_{\bar q} \frac2{\alpha_g}
\left[
\frac{\alpha_q-\alpha_{\bar q}-(2u-1)}{\alpha_g}
[S(\underline{\alpha})+S_\gamma(\underline{\alpha})]+
T_3(\underline{\alpha})-T_2(\underline{\alpha})
\right],\nonumber\\
{\mathbb A}(u)&=&2\int_0^u dv(4u+1-6v)h_\gamma(v)
\\
&&{}-4\int_0^u d\alpha_q\int_0^{\bar u} d\alpha_{\bar q} \frac1{\alpha_g}
\left[
\frac{\alpha_q-\alpha_{\bar q}-(2u-1)}{\alpha_g}\,\left(
T_2(\underline{\alpha})-T_3(\underline{\alpha})\right)-
[S(\underline{\alpha})+S_\gamma(\underline{\alpha})]
\right],\nonumber
\eea
from the first and from the second of the identities in \re{1st}, respectively.
These relations, again,  correspond to (A.13) and (A.14) in \cite{BBKT98} with
the replacement $S\to S+S_\gamma$.
Inserting expressions for the 
three-particle DAs at next-to-leading order in the 
conformal expansion and using (\ref{ferrara}), we obtain for $h_\gamma$:
\bea{eq:h3exp}
h_\gamma(u) & = & -10(1+2\kappa^+)C_2^{1/2}(2u-1).
\eea
The cancellation of contributions $\sim \zeta_i$ is a consequence  
of conformal symmetry. To see this, notice that $h_\gamma$
corresponds to the configuration with both the quark and the antiquark
having ``minus'' light-cone projections and therefore a  
well-defined conformal expansion in terms of Gegenbauer-polynomials 
$C_{j-1}^{1/2}$ where $j$ is the conformal spin, cf.\ \cite{BF90,BBKT98}. 
The symmetry $h_\gamma(u) = h_\gamma(1-u)$ implies that only odd 
conformal spins can contribute, and in addition the contribution
of the lowest possible spin $j=1$ vanishes by EOM. Hence $j=3,5,\dots$,
and in particular the result in \re{eq:h3exp} corresponds to $j=3$.
The $\zeta$-contributions to three-particle DAs have $j=4$; they cannot 
appear in the expansion of $h_\gamma$ and indeed cancel      
by virtue of (\ref{ferrara}).

The situation with $\mathbb A$ is different, since this function
is defined off the light-cone and its conformal expansion is more 
complicated. Using the result in \re{eq:h3exp} and the second 
relation in \re{1sol} we obtain 
\bea{eq:ATexp}
{\mathbb A}(u) & = &
40 u^2 \bar u^2 \left\{
3\kappa-\kappa^+ +1\right\}
 +8 \left(\zeta_2^+ - 3\zeta_2 \right) \times
\nonumber\\
 & & \times \left( u\bar u (2+13 u\bar u) + 2u^3 (10-15u+6u^2) \ln u +
2\bar u^3 (10-15\bar u + 6\bar u^2) \ln \bar u\right).
\hspace*{1cm}\eea

We emphasize that our results for the DAs $h_\gamma$ and $\mathbb A$
are consistent with the approximations adopted for the three-particle
DAs; that is, this set of photon distribution amplitudes 
satisfies all exact QCD identities that are consequence of EOM. 
This issue is nontrivial. For example, consider an important operator 
identity which is an extension of Eq.~(24) in \cite{AB95}:%  
\footnote{
This identity is very useful in practical calculations since it allows 
to express the chiral-odd operator with two free Lorentz-indices through a 
simpler nonlocal operator with one open index and three-particle operators of 
higher twist.} 
\bea{3d}
\bar  q(x) \sigma_{\mu\nu}  q(-x) & = &\int_0^1\!\!\!\! du u
\left\{
\frac{\partial}{\partial x_\nu} \bar  q(ux)\sigma_{\mu x} q(-ux)- (\mu\leftrightarrow \nu)
\right\}-
%\nonumber\\
\varepsilon_{\mu\nu x\xi}\partial_\xi\!\!
\int_0^1\!\!\!\! du u^2\bar  q(ux)\gamma_5 q(-ux)
\nonumber\\
&&-i\varepsilon_{\mu\nu x \xi}\int_0^1\!\!\!\! du u^2\int_{-u}^u\!\!
 dv \bar  q(ux)
\left\{
gG_{x\xi}(vx)+e_q F_{x\xi}(vx)
\right\}\gamma_5 q(-ux)
\nonumber \\
&&+i\int_0^1\!\!\!\! du  u\int_{-u}^u\!\!\!\! dv v \bar  q(ux)
\left\{
[gG_{\mu x}(vx)+e_q F_{\mu x}(vx)]\sigma^{\nu x}-
(\mu\leftrightarrow \nu)\right\}
 q(-ux).\nonumber\\[-10pt]
\eea
Taking the vacuum average in the BG field and using the definitions 
of photon DAs, this implies one more relation   
\bea{check}
{\mathbb A}(u)&=&2\int_0^u dv\,(2v-1)\,h_\gamma(v)
\nonumber \\
&-&4\int_0^u d\alpha_q\int_0^{\bar u} d\alpha_{\bar q} \frac1{\alpha_g}
\left[
\frac{\alpha_q-\alpha_{\bar
q}-(2u-1)}{\alpha_g}\left(T_4(\underline{\alpha})+
T_4^\gamma(\underline{\alpha})-T_3(\underline{\alpha})\right)+
\tilde S (\underline{\alpha})
\right].\nonumber\\[-10pt]\label{3sol}
\eea
One can check that the resulting expression for $\mathbb A$ coincides with 
the result given in \re{eq:ATexp} if and only if full expressions 
for the three-particle DAs and in particular the nonzero result 
for $T_4^\gamma$ \re{Sres} are used. In Ref.~\cite{AB95} the DAs  
$h_\gamma$ and $\mathbb A$ were derived  using
a truncated version of (\ref{3d}) neglecting $T_4^\gamma$ and 
quark-antiquark-gluon operators. In view of relation \re{ferrara} 
this approximation is inconsistent and the resulting DAs 
are superseded by the ones obtained in this paper.

\section{Summary and Conclusions}
\label{sec:5}  
\setcounter{equation}{0} 

In this paper we have given the first comprehensive analysis
of photon distribution amplitudes. 
The special interest (and
complication) springs from the interplay between hard
(perturbative) and soft (nonperturbative) contributions to the photon
structure, as discussed in Sec.~2.2. This interplay becomes nontrivial 
in higher twists, as QCD equations of motion receive additional 
contributions from photon operators. (Alternatively, one may leave 
equations of motion intact, but take into account contact terms.)

We have defined and classified all two- and three-particle
distribution amplitudes of the photon up to twist-4. For the
chiral-odd ones, we could largely draw on the results obtained in
Ref.~\cite{BBK88}, the remaining definitions resemble those for the
$\rho$-meson discussed in \cite{BBKT98,BB98}. The main theoretical  
result of our analysis is the conformal expansion 
of DAs in presence of the background field. We provide a selfconsistent
approximation for photon DAs up to twist-4 that contains a minimal number 
of nonperturbative parameters and is consistent with the exact 
QCD equations of motion. The final expressions  are
collected in Eqs.~(\ref{evenanswer}) (twist-3) and 
 (\ref{Tampl}), (\ref{Sres}), (\ref{eq:h3exp}), (\ref{eq:ATexp}) (twist-4).
The numerical values of the parameters are estimated,
when possible, using QCD sum rules. In particular, we give a new 
estimate of the magnetic susceptibility of the quark condensate 
and derive a sum rule for the normalization of twist-three distributions. 
All parameters are accessible to lattice-simulations,
and we hope that the lattice-community will take up the task and
provide more accurate estimates.

We believe that the results of this work can be applied to the 
studies of hard exclusive processes involving photon emission (or 
absorption) in the framework of QCD factorization. A few examples 
are mentioned in the introduction. A more detailed discussion and concrete 
calculations go beyond the tasks of this paper.

\section*{Acknowledgments}     

The work of N.K.\ is supported by DFG, projects 920585 and 92070. P.B.\ thanks 
the physics department of the university of Regensburg
for kind hospitality during the 
%exhausting 
final stages of this work.

\appendix  
\renewcommand{\theequation}{\Alph{section}.\arabic{equation}}  
\setcounter{table}{0}  
\renewcommand{\thetable}{\Alph{table}}  
   
\section*{Appendices}  

\section{The Conformal Basis}  
\label{app:a}  
\setcounter{equation}{0}  

For a generic three-particle operator a 
``conformal basis'' can be constructed as follows \cite{BDKM,BKM01}:
conformal symmetry allows one to fix the total three-particle conformal
spin $J=j_1+j_2+j_3+N$ of a state.
We define a set of functions ${Y}_{Jj}^{(12)3}$ by requiring
that, in addition to a fixed $J$, they 
also have a definite value of the conformal spin
in the quark-antiquark channel $(12)$, for definiteness:
$j=j_1+j_2+n$ with $n=0,\dots,N$.

Taken together, these two conditions determine the
polynomials ${Y}_{Jj}^{(12)3}$ uniquely and yield the following
expression:
\bea{YJi123}
Y_{Jj}^{(12)3}(\alpha_i) &=&
(1-\alpha_3)^{j-j_1-j_2}\,
  P_{J-j-j_3}^{(2j_3-1,2j-1)}\lr{1-2\alpha_3}
\,P_{j-j_1-j_2}^{(2j_1-1,2j_2-1)}\lr{\frac{\alpha_2-\alpha_1}{1-\alpha_3}}\,.
\eea
Here $P^{(\alpha,\beta)}_n(x)$ is the Jacobi-polynomial.

The basis functions $Y_{Jj}^{(12)3}(\alpha_i)$ are mutually 
orthogonal with respect to the conformal scalar product: 
\beq{scalprod}
%\frac{\Gamma(2j_1+2j_2+2j_3)}
%{\Gamma(2j_1)\Gamma(2j_2)\Gamma(2j_3)}
\int_0^1 {\cal D}\underline{\alpha}\,\,
\alpha_1^{2j_1-1} \alpha_2^{2j_2-1} \alpha_3^{2j_3-1}\,
Y_{Jj}^{(12)3}(\alpha_i) Y_{J'j'}^{(12)3}(\alpha_i) = 
{\cal N}_{Jj}\,\delta_{JJ'}\,\delta_{jj'}\,,
\eeq
where
\bea{normY}
{\cal N}_{Jj}
&=& 
%\frac{\Gamma(2j_1\!+\!2j_2\!+\!2j_3)}
%{\Gamma(2j_1)\Gamma(2j_2)\Gamma(2j_3)}\,
\frac{\Gamma(j\!+\!j_{1}\!-\!j_{2})\Gamma(j\!-\!j_{1}\!+\!j_{2})}
{\Gamma(j\!-\!j_1\!-\!j_2\!+\!1)\Gamma(j\!+\!j_1\!+\!j_2\!-\!1) (2j\!-\!1)}\,
\frac{\Gamma(J\!-\!j\!+\!j_3)\Gamma(J\!+\!j\!-\!j_3)}
{\Gamma(J\!-\!j\!-\!j_3\!+\!1)\Gamma(J\!+\!j\!+\!j_3\!-\!1)(2J\!-\!1)}\,.
\nonumber\\[-5pt]\eea
The above construction of the conformal basis involves the obvious
ambiguity in what  order to couple the spins of partons to the total
spin $J$. Choosing a different two-particle channel
one obtains a  different  conformal basis related to the original one
through the  matrix $\Omega$ of  Racah $6j$-symbols
of SL$(2,\mathbb{R})$, e.g.
\bea{Racah}
{Y}_{Jj}^{(31)2}(x_i)=\sum_{j_1\!+\!j_2\le j'\le J-j_3}\Omega_{jj'}(J)\,
{Y}_{Jj'}^{(12)3}(x_i)\,.
\eea
The properties of the Racah $6j$-symbols as well as 
explicit expressions  in terms of the generalized hypergeometric series
${}_4F_3(1)$ are summarized  in \cite{BKM01}.

\section{\boldmath QCD Sum Rules}
\label{app:b}  
\setcounter{equation}{0}  

Given the considerable time that has passed since the existing numerical
estimates of the parameters of photon DAs 
were obtained \cite{BK84,BKY85,BBK88},
we feel that the corresponding QCD sum rules deserve a fresh look.
Some new results are also included: 
we calculate the radiative correction to the 
leading-twist sum rule and derive a sum rule for the 
normalization of twist-3 distributions. 

\subsection{The leading twist-2 distribution amplitude}

The QCD sum rules for the leading-twist photon DA 
 $\phi_\gamma(u)$ are derived from the correlation 
function of the nonlocal tensor operator on the light-cone 
with the vector current. For massless quarks:
\beq{correlator1}
i\int d^4y \, e^{-iqy} \langle 0 | T [\bar q\gamma_z q](y)
\bar q(0) \sigma_{\mu z}[0,z]q(z)|0\rangle \ = \
i z_\mu\,\frac{qz}{q^2}\qq \int\limits_0^1 du\, e^{-iquz} \Phi(u,q^2) 
\eeq
with 
\beq{correlator2}
 \Phi(u,q^2) = -[\delta(u)+\delta(1-u)] - \frac{\alpha_s C_F}{2\pi} F(u)
 - \frac13 \frac{m_0^2}{q^2}[\delta(u)+\delta(1-u)+\delta'(u)+\delta'(1-u)]
 + {\cal O}(1/q^4)
\eeq 
where $\delta'(x) = (d/dx)\delta(x)$ and 
%\beq{correlator3}
%F(u)  \ = \  2 \left[ \delta(u) + \delta(\bar u)\right] \left(
%\ln\,\frac{-q^2}{\mu^2}-1\right)
% + \left[ \frac{\bar u}{u} \left( 1 -
%\ln\,\frac{-q^2}{\mu^2} - \ln (u\bar u)\right)\right]_+ + \left[ u
%\leftrightarrow \bar u \right]_+ .
%\eeq
\beq{correlator4}
F(u)   =   \left(\ln\,\frac{-q^2}{\mu^2}-1\right)
 \left\{ 2 \delta(u)+2 \delta(\bar u) -
  \left[\frac{\bar u}{u} \right]_+ -\left[\frac{u}{\bar u} \right]_+\right\}
- \left[\frac{\bar u \ln(u\bar u)}{u} \right]_+
- \left[\frac{u \ln(u\bar u)}{\bar u} \right]_+.
\eeq
Here  the $[\ ]_+$ prescription is defined as usual
$$ [f(u)]_+ = f(u) -  \delta(u_0) \int_0^1 dw\,f(w)\,,$$
where $u_0=0$ or $u_0=1$ is the position of the singularity,  
and $m_0^2$ parametrizes the mixed condensate: $\langle \bar q
\sigma gG  q\rangle \equiv m_0^2\qq$. The 
radiative correction to the quark condensate in \re{correlator4} is a 
new result%
\footnote{The expression given in \cite{BB96} for the 
local limit $z\to 0$ contains a typo.}.
By expanding \re{correlator1} around $z=0$
in terms of conformal invariant local operators
one obtains the functions $A_n(q^2)$ given in Eq.~(4.7) in
\cite{BBK88}, up to the terms in $\alpha_s$.

\subsubsection{Magnetic susceptibility} 

The magnetic susceptibility of the quark condensate, $\chi$, is obtained 
from the correlation function 
\beq{chicor}
q^2\chi(q^2) = \int\limits_0^1 \!du\, \Phi(u,q^2) = 
 -2 + \frac{8}{3}\frac{\alpha_s}{\pi}\left(\ln\,\frac{\mu^2}{-q^2}+1\right)
 -\frac{2}{3} \frac{m_0^2}{q^2} + 0 \cdot \frac{\langle g^2 G^2\rangle}{q^4}
 +\ldots,
\eeq 
where we have also indicated that the gluon condensate contribution
vanishes in factorization approximation \cite{BKY85}.
We write the physical spectral density of this  correlation function 
as a sum over several narrow resonances plus a 
smooth continuum starting at a threshold $s_0$. Assuming quark-hadron
duality, the continuum contribution can be represented by the 
perturbative imaginary part of the radiative 
correction, so that
\beq{eq:reso}
\chi(q^2)\qq  = 
- \sum_i \,\frac{m_i f_i f_i^T}{m_i^2-q^2} + \frac{8\alpha_s}{3\pi}\,\qq
\int_{s_0}^\infty ds \,\frac{1}{s(s-q^2)},
\eeq  
where $f_i$ and $f_i^T$ denote the couplings of the resonance $i$ to the
vector and tensor current, and are defined as in \re{f-fT}.
The factor $8/3$ in front of the integral is related to the anomalous 
dimensions defined in \re{scale}:  
$1/2 (\gamma_0 - \gamma_{\qq}) = 8/3$.
Once these parameters are fixed, the magnetic susceptibility is obtained 
from \re{eq:reso} as $\chi = \chi(0)$.

In a first approximation one can retain the contribution of the 
lowest-lying $\rho$-meson state only and use the same value of the continuum
threshold as obtained in the sum rules for the correlation function 
of vector currents, $s_0= 1.5\,$GeV$^2$ \cite{SVZ}. Using the  
QCD sum rule estimate
$f_\rho^\perp = 160\,$MeV one obtains
\beq{VDMimproved}
     \chi(1\,\mbox{GeV}) = -\frac{f_\rho f^\perp_\rho}{m_\rho\qq} + 
      \frac{8\alpha_s}{3\pi}\frac{1}{s_0} \simeq
       3.0\,\mbox{GeV}^{-2}\,, 
\eeq
which is an improvement over the pure VDM estimate in \re{VDM2}.
Including more resonances and at the same time increasing the continuum 
threshold, one may hope to get a better accuracy.  
The corresponding sum rules can be constructed in two different ways: either  
matching the expansion of \re{eq:reso} for $q^2\to -\infty$ to the 
power series in \re{chicor} \cite{BKY85} or performing 
the Borel-transformation 
of both expressions and matching the results for intermediate values 
of the Borel-parameter of the order of a few GeV \cite{BK84}.
Once the parameters are fixed, the magnetic susceptibility is obtained 
from \re{eq:reso} as $\chi = \chi(0)$.
 
\begin{figure}
$$\epsfysize=4cm\epsffile{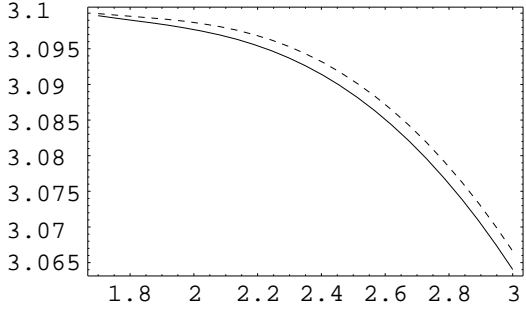}$$
\caption[]{$\chi(1\,\mbox{GeV})$ as function of $m_4$ with $s_0$ fixed as
$s_0 = (m_4 + 0.1\,\mbox{GeV})^2$ (solid line) and $(m_4 +
0.2\,\mbox{GeV})^2$ (dashed line).}\label{fig:app1}
\end{figure}

Following the first procedure we obtain the so-called local-duality sum rules:
\bea{eq:grunz}
\sum m_i f_i f_i^T(\mu) &= & -2 \qq(\mu) \left[1 + \frac{4\alpha_s(\mu)}{3\pi}
\left( \ln\,\frac{s_0}{\mu^2} - 1 \right)\right],\nonumber\\
\sum m_i^3 f_i f_i^T(\mu) &= & -2 \qq(\mu) \left[\frac{4\alpha_s(\mu)}{3\pi}
s_0 + \frac{m_0^2(\mu)}{3}\right],\nonumber\\
\sum m_i^5 f_i f_i^T(\mu) &= & -4 \qq(\mu) \left[\frac{\alpha_s(\mu)}{3\pi}
s_0^2 + 0 \cdot \langle G^2\rangle \right].
\eea
For the couplings of the ground-state resonance $\rho$(770) one  can  use the 
values quoted in \re{f-fT}.
The three local-duality sum rules in \re{eq:grunz} can be 
used to determine the contributions of three more resonances, so that we
truncate the sums in \re{eq:grunz} after four terms.
Experimentally,
the three lowest-lying resonances $\rho(770)$, $\rho(1450)$ and 
$\rho(1700)$ are established \cite{PDG2000}. 
The magnetic susceptibility $\chi(\mu)$ then exhibits a
mild dependence on $s_0$ and on the unknown mass of the fourth resonance 
$m_4$. As an example,
we show in Fig.~\ref{fig:app1} $\chi(1\,\mbox{GeV})$  as function of 
$m_4$ with $s_0$ fixed as
$s_0 = (m_4 + 0.1\,\mbox{GeV})^2$ (solid line) and $(m_4 +
0.2\,\mbox{GeV})^2$ (dashed line). In this calculation we use  
$\qq(1\,\mbox{GeV}) = (-0.25\,\mbox{GeV})^3$, $m_0^2(1\,\mbox{GeV}) =
0.65\,$GeV$^2$, cf.\ \cite{BB96},  and the value 
$\alpha_s(1\,\mbox{GeV}) = 0.51$.

Neglecting for the moment the
uncertainty of $f_1 f_1^T$ and the local-duality approximation itself, 
we obtain from local-duality sum rules
with four resonances
\beq{chilocal}
\chi(1\,\mbox{GeV}) = (3.08\pm 0.02)\,\mbox{GeV}^{-2}.
\eeq
For a typical value of $m_4$, the breakdown of the total number in
contributions from the different resonances and the continuum is
$2.73 +0.30 - 0.07 +0.04 + 0.10$, which shows a clear dominance of
$\rho(770)$.
Neglecting radiative corrections altogether, we would find $\chi = (3.3\pm
0.1)\,\mbox{GeV}^{-2}$. 

As a consistency check, we relax the condition that the known value of
$f_1 f_1^T$ be used as input and include only three resonances
in the sums in (\ref{eq:grunz}). We then find $f_1^T(1\,\mbox{GeV}) = (170\pm
7)\,\mbox{MeV}$ for $m_3^2\leq s_0\leq 5\,\mbox{GeV}^2$, which is
consistent with the result from Borel sum-rules quoted above, and 
$\chi(1\,\mbox{GeV}) = (3.20\pm 0.15)\,\mbox{GeV}^{-2}$, in complete
agreement with (\ref{chilocal}). 

\begin{figure}
$$
\epsfysize=4cm\epsffile{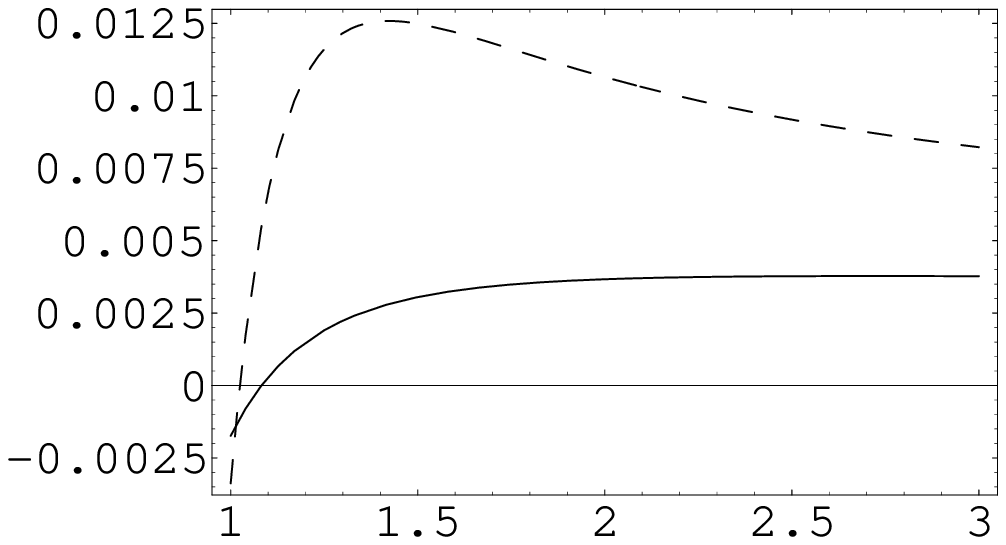}
\quad
\epsfysize=4cm\epsffile{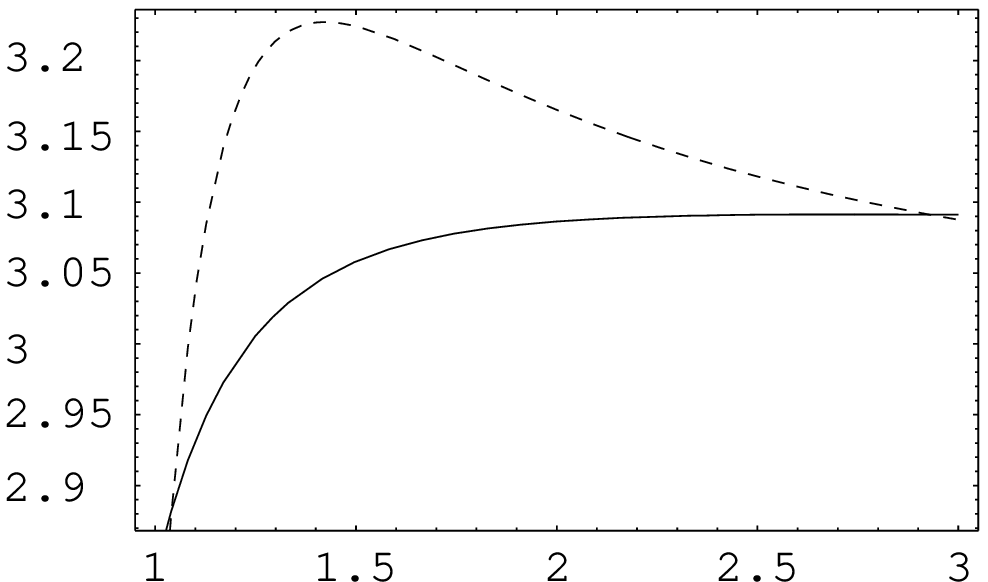}
$$
\caption[]{(a) Coupling $f_{\rho'}f_{\rho'}^T$ from the Borel sum-rule
\protect{(\ref{eq:appBSR})} as a function of the 
Borel-parameter. Solid line: $m_{\rho'}=1.4\,$GeV, dashed line:
$m_{\rho'}=2\,$GeV. $s_0 = (m_{\rho'}+0.1\,\mbox{GeV})^2$. (b)
$\chi(1\,\mbox{GeV})$ from \protect{(\ref{eq:reso})} with $q^2=0$ with
two resonances, $\rho(770)$ and an effective $\rho'$, as a function of
the Borel-parameter. Identification
of curves as in (a).}\label{fig:app2}
\end{figure}

Alternatively, and this constitutes the 2nd procedure, one
 investigates the Borel sum-rule
\bea{eq:appBSR}
\lefteqn{
m_\rho f_\rho f_\rho^T e^{-m_\rho^2/M^2} + m_{\rho'} f_{\rho'} f_{\rho'}^T 
e^{-m_{\rho'}^2/M^2} =}\hspace*{0.5cm}\nonumber\\
& = & -2\qq \left[ 1 +
\frac{4}{3}\,\frac{\alpha_s}{\pi} \left( \ln\,\frac{M^2}{\mu^2} -
\gamma_E - 1 - \int_{s_0}^\infty \frac{ds}{s}\, e^{-s/M^2}\right) -
\frac{1}{3}\,\frac{m_0^2}{M^2} + 0\cdot \frac{\langle
G^2\rangle}{M^4}\right],\hspace*{0.5cm}
\eea
where for simplicity we have substituted the sum over higher resonances 
by the contribution of one effective state $\rho'$ and consider 
the  mass $m_{\rho'}$ as a free parameter. The sum rule \re{eq:appBSR}
can be used to extract the product $f_{\rho'}
f_{\rho'}^T$ as a function of $M^2$, $s_0$ and $m_{\rho'}$. We find
that for $m_{\rho'}\geq 1.4\,$GeV there is a reasonable 
stability-plateau in the Borel-parameter $M^2$, as shown in 
Fig.~\ref{fig:app2}.
Varying $m_{\rho'}$ between 1.4 and
2.0~GeV and fixing the continuum threshold as $s_0 =
(m_{\rho'}+0.1\,\mbox{GeV})^2$, we find from (\ref{eq:reso})
\beq{eq:chiBorel}
\chi(1\,\mbox{GeV}) = (3.15\pm 0.10)\,\mbox{GeV}^{-2},
\eeq
which is perfectly consistent with the value obtained from local-duality 
sum-rules. 

We are now in a position to quote our final value for
$\chi(1\,\mbox{GeV})$. Averaging over the results from local-duality
sum rules with three and four resonances, and the result from Borel 
sum-rules, we obtain the central value $3.15\,$GeV$^{-2}$. In estimating
the uncertainty, we take into account that the $\rho(770)$ is the
dominant contribution, so that $\Delta \chi/\chi = \Delta(f_\rho
f_\rho^T)/(f_\rho f_\rho^T)$ in good approximation. Adding in
quadrature the uncertainties from $\Delta
f_\rho = 10\,$MeV, $\Delta
f_\rho^T = 10\,$MeV and the uncertainty quoted in (\ref{eq:chiBorel}),
we finally arrive at the estimate quoted in \re{SR-chi-new}.
%\begin{equation}\label{eq:chifinal}
%\chi(1\,\mbox{GeV}) = -(3.15\pm 0.30)\,\mbox{GeV}^{-2}.
%\end{equation}

\subsubsection{Higher moments}

Let us now turn to the shape of the photon DA.
 At first sight, assuming
the vector-dominance approximation in \re{correlator1} $1/(-q^2)\to 1/m_\rho^2$
at $q^2\to 0$, one is led to the conclusion that the photon DA
 schould be strongly peaked at the end-points, with the 
quark condensate contribution $\delta(u)+\delta(1-u)$ somewhat smoothened 
by contributions of condensates of higher dimension. We start by giving  
a qualitative argument why this picture is wrong.

To this end, it is convenient to consider the Gegenbauer-moments 
of the correlation function \re{correlator1} defined as
\bea{Gegmoments}
   \chi_n(q^2) &=& \frac{1}{q^2}
   \frac{2(2n+3)}{3(n+1)(n+2)}\int_0^1 du\, C^{3/2}_n(2u-1) \Phi(u,q^2)
 \nonumber\\
   &=& -\frac{2(2n+3)}{3}\left[\frac{1}{q^2}+ \frac{(n+1)(n+2)}{6}
   \frac{m_0^2}{q^4} + {\cal O}(n^4/q^6)\right],
\eea
where for simplicity we suppress radiative corrections.
Obviously $\chi_0(q^2) = \chi(q^2)$, and the coefficients $\phi_n$ 
in the expansion of the photon DA over the Gegenbauer-polynomials 
\re{scale} are given by $\phi_n = \chi_n(q^2=0)/\chi$. 
Assuming that the correlation function $\chi_n$ for arbitrary $n$ 
is saturated by the contribution of the $\rho$-meson and using 
\re{locdual}, one obtains 
\beq{garbage1}
     \phi^{\rm VDM}_n = \frac{(2n+3)}{3} 
\eeq
and 
\beq{garbage2}
  \phi_\gamma(u) = 6 u \bar u \sum_{n\,\rm{even}}
\frac{1}{3}\, (2n+3)\, C_n^{3/2}(2u-1) = 
\frac12\Big[\delta(u) + \delta(\bar u)\Big]\,, 
\eeq 
as expected.

To see why this is wrong, consider the expansion of the VDM-type 
contribution in a power series 
\beq{VDMn}
 \chi_n(q^2) \simeq \frac{1}{m^2-q^2} = -\frac{1}{q^2}- \frac{m^2}{q^4} +\ldots
\eeq 
The comparison of the two expansions \re{Gegmoments} and \re{VDMn}
shows that in order for them to be consistent the  mass of the resonance has 
to grow rapidly with $n$: $m_n \sim n$. Alternatively, one can say that 
the low-mass $\rho$-meson constribution becomes irrelevant for large $n$.

To see what this means, let us generalize the vector-dominance 
approximation by 
allowing the effective mass of the resonance to be $n$-dependent and, in 
particular, choose the mass such as to reproduce the first two terms 
in the operator product expansion:
\beq{model1}
    m^2_{n,{\rm eff}} = m_0^2(n+1)(n+2)/6\,.
\eeq  
It is easy to see that in this approximation 
\beq{model2}
   \phi_n = \frac{(2n+3)}{3} \frac{m^2_{0,{\rm eff}}}{m^2_{n,{\rm eff}}}
 = \frac{2(2n+3)}{3(n+1)(n+2)}\,.
\eeq  
One obtains in this model
\beq{model3}
   \phi_\gamma(u) = 6 u \bar u \sum_{n\,\rm{even}}
\frac{2(2n+3)}{3(n+1)(n+2)}\, C_n^{3/2}(2u-1) = 1 \,, 
\eeq
that is a flat DA with no momentum fraction 
dependence: the peaks at the end-points have disappeared.
Note that our argumentation only invokes the structure of nonperturbative
power-like corrections in the operator product expansion. Taking into account  
perturbative gluon radiation is expected to further suppress the end-point 
behaviour, as illustrated in Fig.~\ref{fig:phigamma} for the example of the 
instanton model.    

Trying to be somewhat more quantitative, we can write the generalization of 
(\ref{eq:reso}) to arbitrary Gegenbauer-moments:
\begin{equation}
\chi_n(q^2) \qq = -\sum_i\,\frac{m_i f_i
f_i^T\phi_{i,n}}{m_i^2-q^2} +
\frac{2}{3}\,(2n+3)\left(\sum_{j=1}^{n+1}\frac{1}{j}\right) 
\frac{\alpha_s}{\pi}\,C_F\,\qq \int_{s_0}^\infty
ds\,\frac{1}{s(s-q^2)},
\end{equation}
where $\phi_{i,n}$ is the $n$th coefficient in the Gegenbauer-expansion of the 
DA of the $i$th
resonance. Note that $\gamma_n - \gamma_{\qq} = 4 C_F
\sum_{j=1}^{n+1}(1/j)$.
For $n=2$, one can construct local-duality sum rules:
\bea{eq:grunz2}
\sum m_i f_i f_i^T(\mu) \phi_{i,2}(\mu) 
&= & -\frac{14}{3}\, \qq(\mu) \left[1 + 
\frac{4}{3}\,\frac{\alpha_s(\mu)}{\pi}
\left( \frac{11}{6}\,\ln\,\frac{s_0}{\mu^2} - \frac{131}{36} 
\right)\right],\nonumber\\
\sum m_i^3 f_i f_i^T(\mu) \phi_{i,2}(\mu) &= & -\frac{14}{3}\, \qq(\mu) 
\left[\frac{22}{9}\,\frac{\alpha_s(\mu)}{\pi}
s_0 + 2 m_0^2(\mu)\right],
%\nonumber\\
%\sum m_i^5 f_i f_i^T(\mu) \phi_{i,2}(\mu) 
%&= & -\frac{14}{3}\, \qq(\mu) \left[\frac{11}{9}\,\frac{\alpha_s(\mu)}{\pi}
%s_0^2 + \frac{85}{108} \langle g^2G^2\rangle \right].
\eea
where we suppress a third sum rule, analogous to the 3rd equation in
\re{eq:grunz}, 
because the factorization approximation for the gluon condensate 
contribution is known to be unreliable.

A comparison with (\ref{eq:grunz}) shows that the expression
on the r.-h.s.\ of the first sum rule is by
a factor 1.36 larger (assuming $s_0 = 2.5\,$GeV$^2$),
 whereas the r.-h.s\ of the second sum rule 
increases by roughly a factor 3. This is consistent with the pattern 
suggested above, that higher resonances start to play a 
decisive r\^ole. Using a model with two resonances, $\rho(770)$ and 
$\rho(1450)$, and the continuum treshold $s_0= 2.5\,$GeV$^2$ one obtains 
the small value $\phi_2 = 0.07$. The problem with this procedure is that 
the fit ``wants''  a negative value of the second Gegenbauer-coefficient
$\phi_{\rho,2}$, which contradicts a more direct QCD sum rule evaluation 
of this coefficient from diagonal sum-rules \cite{BB96}. If the
value \cite{BB96}, $\phi_{\rho,2} = 0.2\pm 0.1$, is enforced, and, say,
the mass of the resonance is taken as free parameter, the result for 
$\phi_2$ increases by almost an order of magnitude. In our point of view 
such an instability  indicates that no reliable conclusion can be drawn 
on the value of $\phi_2$ in the QCD sum rule approach.   
  
\subsection{Twist-3 Distributions}

To leading-order accuracy of the conformal expansion 
we only need the parameter $f_{3\gamma}$ defined
in \re{def:f3}. It can be estimated from the correlation 
function
\beq{CF}
i\!\!\int\!\! d^4 x e^{-iqx}\langle 0| 
 T \bar q(0) \tilde G_{z \nu}(0)\gamma_z\gamma_5 q(0)
\bar q (x)\gamma_\beta q (x)
|0\rangle = 
 f_{3\gamma}(q^2) (qz) (z_{\beta}q_\nu-(qz)g_{\nu\beta})+ \dots,
\eeq
where the Lorentz-structure is chosen in order to 
pick up the relevant coupling of twist-3 corresponding to the 
transverse projection $G_{z\nu} \to G_{z\perp}$ and to
suppress contributions of twist-4. 
%dots denote irrelevant structures of higher twist. 
$f_{3\gamma}$ is obtained as 
$f_{3\gamma}(0)$.
The operator product expansion of $f_{3\gamma}(q^2)$, up to
contributions of dimension 6, reads
\bea{cond}
f_{3\gamma}(q^2) & = &
\frac{\alpha_s}{4\pi}\,\frac{q^2}{\pi^2}\left(\frac{1}{36}\,\ln\,
\frac{-q^2}{\mu^2}  + \mbox{const.}\right) +
\frac{1}{24q^2}\,\left\langle
\frac{\alpha_s}{\pi}\, G^2\right\rangle - \frac89 \, 
\frac{\pi\alpha_s\qq^2}{q^4}\,,
\eea
and the $\rho$-meson contribution can be obtained by 
a standard QCD sum-rule calculation,
\bea{shit2}
m_\rho^2 f_\rho^2 \zeta_{3\rho} e^{-m_\rho^2/M^2} & = &
\frac{\alpha_s}{4\pi}\,\frac{1}{36\pi^2} \int_0^{s_0} ds\, s
e^{-s/M^2} + \frac{1}{24}\,\left\langle
\frac{\alpha_s}{\pi}\, G^2\right\rangle + \frac89 \, 
\frac{\pi\alpha_s\qq^2}{M^2}.
\eea 
Using $s_0=1.5$~GeV$^2$ and 
$\langle\frac{\alpha_s}{\pi}\, G^2\rangle=0.012$~GeV$^4$~\cite{SVZ},
we obtain from this sum rule (at the normalization point $\mu=1\,$GeV)
\beq{zeta}
\zeta_{3\rho}(1\,{\rm GeV})= 0.10\pm 0.05,
\eeq 
which is by a factor 3 larger than the result quoted in \cite{BBKT98}
and obtained in Ref.~\cite{ZZC} 
from the diagonal correlation function of two quark-antiquark-gluon
operators.    

In vector-dominance approximation $f_{3\gamma} = -f_\rho^2 \zeta_{3\rho}$. 
Including the continuum
contributions, $f_{3\gamma}$ is obtained as\footnote{In order to
obtain this formula, we express the continuum contribution to
(\ref{cond}) via a dispersion relation 
$\sim \int_{s_0}^{\mu_F^2} ds = \mu_F^2-s_0$,
where $\mu_F$ is the factorization scale.  The quadratically 
divergent term $\sim \mu_F^2$ matches the contribution of low momenta in the 
perturbative contributions of twist-1 in \re{OPEeven} which have to 
be subtracted to avoid double counting. We prefer to define perturbative 
contributions in such a way as to 
include all momenta down to zero, and therefore 
have to omit quadratically divergent parts of the higher-twist operators.
Thus we only keep the term in $s_0$. 
%dispersion relation refers to a hard-cutoff scheme, whereas in
%Sec.~2.2 we have implicititly assumed the $\overline{\rm MS}$ scheme,
%in which power-like divergencies $\sim \mu_F^2$ are included in the
%hard perturbative contribution. In the sum-rule for $f_{3\gamma}$ we
%thus assume that this term has already been accounted for and only
%keep the term in $\sim s_0$.
}
\beq{shit1}
f_{3\gamma}(0)=-f_\rho^2 \zeta_{3\rho} +
\frac{\alpha_s}{4\pi}\,\frac{s_0}{\pi^2}\frac{1}{36}.
\eeq 
Using the value of $\zeta_{3\rho}$ given in \re{zeta} 
we obtain the estimate (at the normalization point $\mu=1\,$GeV)
\beq{fest}
f_{3\gamma}=-(0.0039\pm 0.0020)\,\mbox{GeV}^2
\eeq
with $s_0=1.5\,$GeV$^2$. The correction to the vector-dominance approximation
is small, $\sim5\,$\%.

To next-to-leading order in the conformal spin, we have two
  parameters,  $\omega^A_\gamma$ and  $\omega^V_\gamma$, which  
are defined by the following matrix-elements:
\bea{eq:4.7}
\langle 0|\bar  q \zslash (gG_{\alpha\nu}
z^\alpha (i\derright z) - (i\derleft z) g G_{\alpha\nu}z^\alpha)
 q |0\rangle_F  
 &=& e_q (pz)^2 F_{\nu z} 
f_{3\gamma}  \frac{3}{28} \,\omega^V_\gamma + O(z_\nu),
\nonumber\\
\langle 0|\bar  q \zslash\gamma_5 \left[ i Dz,g\tilde{G}_{\mu\nu}
  z^\mu \right]  q -\frac{3}{7} (i\partial z)
\bar  q \zslash\gamma_5 g\tilde{G}_{\mu\nu} z^\mu  q
|0\rangle_F &=& e_q i(pz)^2  F_{\nu z} f_{3\gamma}  \frac{3}{28}\,
\omega^A_\gamma + O(z_\nu),\nonumber\\[-10pt]
\eea
where $[.,.]$ stands for the commutator.
In the vector-dominance approximation
$\omega^{A}_\gamma = \omega^A_{3\rho}$ and 
$\omega^{V}_\gamma = \omega^V_{3\rho}$.
The numbers given in \re{VDMtwist3} are quoted from \cite{ZZC} 
and have to be taken with caution.  
Since even for leading twist we have been not able to draw reliable 
conclusions on similar parameters from the sum rules, we
consider the given numbers as an order-of-magnitude estimate.
From the general experience of similar calculations we expect that 
this is rather an upper bound.


\addcontentsline{toc}{section}{References}
\begin{thebibliography}{99}

\bibitem{massengrab2}
P.\ Kroll and M.\ Raulfs,
%``The pi gamma transition form factor and the pion wave function,''
Phys.\ Lett.\ B {\bf 387} (1996) 848;\\
%%CITATION = HEP-PH 9605264;%%
A.\ V.\ Radyushkin and R.\ T.\ Ruskov,
%``Transition Form Factor gamma gamma* $\to$ pi0 and QCD sum rules,''
Nucl.\ Phys.\ B {\bf 481} (1996) 625;\\
%%CITATION = HEP-PH 9603408;%%
I.\ V.\ Musatov and A.\ V.\ Radyushkin,
%``Transverse momentum and Sudakov effects in exclusive QCD processes:  
%gamma* gamma pi0 form factor,''
Phys.\ Rev.\ D {\bf 56} (1997) 2713.
%%CITATION = HEP-PH 9702443;%%

\bibitem{massengrab3}
D.\ M\"uller {\it et al.},
%``Wave functions, evolution equations and evolution kernels from light-ray operators of {QCD},''
Fortsch.\ Phys.\  {\bf 42} (1994) 101;\\
%[arXiv:hep-ph/9812448];\\
%%CITATION = HEP-PH 9812448;%%
X.\ D.\ Ji,
%``Gauge invariant decomposition of nucleon spin,''
Phys.\ Rev.\ Lett.\  {\bf 78} (1997) 610;\\
%%CITATION = HEP-PH 9603249;%%
A.\ V.\ Radyushkin,
%``Scaling Limit of Deeply Virtual Compton Scattering,''
Phys.\ Lett.\ B {\bf 380} (1996) 417.
%%CITATION = HEP-PH 9604317;%%

\bibitem{massengrab1}
M.\ Beneke, T.\ Feldmann and D.\ Seidel,
%``Systematic approach to exclusive B $\to$ V l+ l-, V gamma decays,''
Nucl.\ Phys.\ B {\bf 612} (2001) 25;\\
%%CITATION = HEP-PH 0106067;%%
S.\ W.\ Bosch and G.\ Buchalla,
%``The radiative decays B $\to$ V gamma at next-to-leading order in QCD,''
Nucl.\ Phys.\ B {\bf 621} (2002) 459;\\
%%CITATION = HEP-PH 0106081;%%
A.\ Ali and A.\ Y.\ Parkhomenko,
%``Branching ratios for B $\to$ K* gamma and B $\to$ rho gamma decays 
%in next-to-leading order in the large energy effective theory,''
Eur.\ Phys.\ J.\ C {\bf 23} (2002) 89.
%%CITATION = HEP-PH 0105302;%%

\bibitem{photonSF}
E.\ Witten,
%``Anomalous Cross-Section For Photon - Photon Scattering In Gauge Theories,''
Nucl.\ Phys.\ B {\bf 120} (1977) 189;\\
%%CITATION = NUPHA,B120,189;%%
W.\ A.\ Bardeen and A.\ J.\ Buras,
%``Higher Order Asymptotic Freedom Corrections To Photon - Photon Scattering,''
Phys.\ Rev.\ D {\bf 20} (1979) 166
[E: D {\bf 21} (1979) 2041];\\
%%CITATION = PHRVA,D20,166;%%
D.\ W.\ Duke and J.\ F.\ Owens,
%``The Photon Structure Function As Calculated Using Perturbative 
%Quantum Chromodynamics,''
Phys.\ Rev.\ D {\bf 22} (1980) 2280.
%%CITATION = PHRVA,D22,2280;%%

\bibitem{Schwinger}
J.\ S.\ Schwinger,
%``On Gauge Invariance And Vacuum Polarization,''
Phys.\ Rev.\  {\bf 82} (1951) 664.
%%CITATION = PHRVA,82,664;%%

\bibitem{Abbott80}
L.\ F.\ Abbott,
%``The BG Field Method Beyond One Loop,''
Nucl.\ Phys.\ B {\bf 185} (1981) 189;
%%CITATION = NUPHA,B185,189;%%
%``Introduction To The BG Field Method,''
Acta Phys.\ Polon.\ B {\bf 13} (1982) 33.
%%CITATION = APPOA,B13,33;%%

\bibitem{SV81}
E.\ V.\ Shuryak and A.\ I.\ Vainshtein,
%``Theory Of Power Corrections To Deep Inelastic Scattering In Quantum 
% Chromodynamics. 1. Q**2 Effects,''
Nucl.\ Phys.\ B {\bf 199} (1982) 451.
%%CITATION = NUPHA,B199,451;%%

\bibitem{DS81}
M.\ S.\ Dubovikov and A.\ V.\ Smilga,
%``Analytical Properties Of The Quark Polarization Operator In An External Selfdual Field,''
Nucl.\ Phys.\ B {\bf 185} (1981) 109.
%%CITATION = NUPHA,B185,109;%%

\bibitem{IS84}
B.\ L.\ Ioffe and A.\ V.\ Smilga,
%``Nucleon Magnetic Moments And Magnetic Properties Of Vacuum In QCD,''
Nucl.\ Phys.\ B {\bf 232} (1984) 109.
%%CITATION = NUPHA,B232,109;%%


\bibitem{NSVZ85}
V.\ A.\ Novikov {\it et al.}, 
%``Calculations In External Fields In Quantum Chromodynamics:. 
%Technical Review (Abstract Operator Method, Fock-Schwinger Gauge),''
Fortsch.\ Phys.\  {\bf 32} (1985) 585.
%%CITATION = FPYKA,32,585;%%

\bibitem{BBK88}
I.\ I.\ Balitsky, V.\ M.\ Braun and A.\ V.\ Kolesnichenko,
%``The Decay Sigma+ $\to$ P Gamma In QCD: Bilocal Corrections In A 
%Variable Magnetic Field And The Photon Wave Functions,''
Sov.\ J.\ Nucl.\ Phys.\  {\bf 48} (1988) 348
[Yad.\ Fiz.\  {\bf 48} (1988) 547];
%%CITATION = SJNCA,48,348;%%
%I.\ I.\ Balitsky, V.\ M.\ Braun and A.\ V.\ Kolesnichenko,
%``Radiative Decay Sigma+ $\to$ P Gamma In Quantum Chromodynamics,''
Nucl.\ Phys.\ B {\bf 312} (1989) 509.
%%CITATION = NUPHA,B312,509;%%

\bibitem{earlyCZ}
 V.\ L.\ Chernyak and A.\ R.\ Zhitnitsky, JETP Lett.\ {\bf {25}} (1977) 510;  
Sov.\ J.\ Nucl.\ Phys.\  {\bf 31} (1980) 544
[Yad.\ Fiz.\  {\bf 31} (1980) 1053];\\
%%CITATION = JTPLA,25,510;%%
%%CITATION = SJNCA,31,544;%%
 V.\ L.\ Chernyak, V.\ G.\ Serbo and A.\ R.\ Zhitnitsky, JETP Lett.\ {\bf  
 26} (1977) 594; Sov.\ J.\ Nucl.\ Phys.\  {\bf 31} (1980) 552
[Yad.\ Fiz.\  {\bf 31} (1980) 1069];\\
%%CITATION = SJNCA,31,552;%%
%\bibitem{earlyBL}
  G.\ P.\ Lepage and S.\ J.\ Brodsky, Phys.\ Lett.\ B {\bf 87} (1979) 359;  
  Phys.\ Rev.\ Lett.\ {\bf 43} (1979) 545 [E: {\bf 43} (1979) 
  1625];
%%CITATION = PHLTA,B87,359;%%
%%CITATION = PRLTA,43,545;%%
  Phys.\ Rev.\ D {\bf 22} (1980) 2157;  \\
   S.\ J.\ Brodsky, G.\ P.\ Lepage and A.\ A.\ Zaidi,  
  Phys.\ Rev.\ D  {\bf 23} (1981) 1152;\\
%%CITATION = PHRVA,D23,1152;%%
%\bibitem{earlyER}
  A.\ V.\ Efremov and A.\ V.\ Radyushkin, Phys.\ Lett.\ B {\bf 94} (1980)  
%%CITATION = PHLTA,B94,245;%%
  245; 
Theor.\ Math.\ Phys.\  {\bf 42} (1980) 97
[Teor.\ Mat.\ Fiz.\  {\bf 42} (1980) 147].
%%CITATION = TMPHA,42,97;%%


\bibitem{AB95}
A.\ Ali and V.\ M.\ Braun,
%``Estimates of the weak annihilation contributions to the decays 
%B $\to$ rho + gamma and B $\to$ omega + gamma,''
Phys.\ Lett.\ B {\bf 359} (1995) 223.
%[arXiv:hep-ph/9506248].
%%CITATION = HEP-PH 9506248;%%

\bibitem{PPRWG98}
V.\ Y.\ Petrov {\it et al.},
%\ V.\ Polyakov, R.\ Ruskov, C.\ Weiss and K.\ Goeke,
%``Pion and photon light-cone wave functions from the instanton vacuum,''
Phys.\ Rev.\ D {\bf 59} (1999) 114018;\\
%%CITATION = HEP-PH 9807229;%%
M.\ Praszalowicz and A.\ Rostworowski,
%``Pion light cone wave function in the non-
local NJL model,''
Phys.\ Rev.\ D {\bf 64} (2001) 074003.
%%CITATION = HEP-PH 0105188;%%

\bibitem{BB89}
I.\ I.\ Balitsky and V.\ M.\ Braun,
%``Evolution Equations For QCD String Operators,''
Nucl.\ Phys.\ B {\bf 311} (1989) 541.
%%CITATION = NUPHA,B311,541;%%

\bibitem{BBKR94}
V.\ M.\ Belyaev {\it et al.},
%``D* D pi and B* B pi couplings in QCD,''
Phys.\ Rev.\ D {\bf 51} (1995) 6177.
%%CITATION = HEP-PH 9410280;%%

\bibitem{Collins}
J.\ C.\ Collins, {\it Renormalization}, 
%``Renormalization. An Introduction To Renormalization, 
%The Renormalization Group, And The Operator Product Expansion,''
Cambridge University Press, Cambridge (1984).

\bibitem{BB91}
I.\ I.\ Balitsky and V.\ M.\ Braun,
%``The Nonlocal Operator Expansion For Inclusive Particle Production 
%In E+ E- Annihilation,''
Nucl.\ Phys.\ B {\bf 361} (1991) 93.
%%CITATION = NUPHA,B361,93;%%

\bibitem{Book}
J.\ D.\ Bjorken and S.\ D.\ Drell,
{\it Relativistic Quantum Mechanics}, 
McGraw-Hill, New York (1964).


\bibitem{Wust97}
M.\ W\"usthoff,
%``Large rapidity gap events in deep inelastic scattering,''
Phys.\ Rev.\ D {\bf 56} (1997) 4311.
%[arXiv:hep-ph/9702201].
%%CITATION = HEP-PH 9702201;%%

\bibitem{IW99}
D.\ Y.\ Ivanov and M.\ W\"usthoff,
%``Hard diffractive photon proton scattering at large t,''
Eur.\ Phys.\ J.\ C {\bf 8} (1999) 107.
%[arXiv:hep-ph/9808455].
%%CITATION = HEP-PH 9808455;%%

\bibitem{CZreport}
V.\ L.\ Chernyak and A.\ R.\ Zhitnitsky,
%``Asymptotic Behavior Of Exclusive Processes In QCD,''
Phys.\ Rept.\  {\bf 112} (1984) 173.
%%CITATION = PRPLC,112,173;%%

\bibitem{BB96}
P.\ Ball and V.\ M.\ Braun,
%``The $\rho$ Meson Light-Cone Distribution Amplitudes of Leading Twist 
%Revisited,''
Phys.\ Rev.\ D {\bf 54} (1996) 2182.
%%CITATION = HEP-PH 9602323;%%

\bibitem{SVy81}
M.\ A.\ Shifman and M.\ I.\ Vysotsky,
%``Form-Factors Of Heavy Mesons In QCD,''
Nucl.\ Phys.\ B {\bf 186} (1981) 475.
%%CITATION = NUPHA,B186,475;%%


\bibitem{BY83}
I.\ I.\ Balitsky and A.\ V.\ Yung,
%``Proton And Neutron Magnetic Moments From QCD Sum Rules,''
Phys.\ Lett.\ B {\bf 129} (1983) 328.
%%CITATION = PHLTA,B129,328;%%

\bibitem{PDG2000}
D.\ E.\ Groom {\it et al.}  [Particle Data Group],
%``Review Of Particle Physics,''
Eur.\ Phys.\ J.\ C {\bf 15} (2000) 1.
%%CITATION = EPHJA,C15,1;%%

\bibitem{BK84}     
V.\ M.\ Belyaev and Y.\ I.\ Kogan,
%``Calculation Of Quark Condensate Magnetic Susceptibility By QCD Sum 
%Rule Method,''
Yad.\ Fiz.\  {\bf 40} (1984) 1035.
%%CITATION = YAFIA,40,1035;%%

\bibitem{BKY85}     
I.\ I.\ Balitsky, A.\ V.\ Kolesnichenko and A.\ V.\ Yung,
%``On Vector Dominance In Sum Rules For Electromagnetic Hadron 
%Characteristics. (In Russian),''
Yad.\ Fiz.\  {\bf 41} (1985) 282.
%%CITATION = YAFIA,41,282;%%

\bibitem{photondijets}
V.\ M.\ Braun {\it et al.}, 
%``Exclusive photoproduction of hard dijets and 
%magnetic susceptibility of  QCD vacuum,''
hep-ph/0206305.
%%CITATION = HEP-PH 0206305;%%

\bibitem{SVZ}
M.\ A.\ Shifman, A.\ I.\ Vainshtein and V.\ I.\ Zakharov,
%``QCD And Resonance Physics. Sum Rules,''
Nucl.\ Phys.\ B {\bf 147} (1979) 385, 448.
%%CITATION = NUPHA,B147,385;%%
%%CITATION = NUPHA,B147,448;%%

\bibitem{BBKT98}
P.\ Ball {\it et al.},
%``Higher twist distribution amplitudes of vector mesons in {QCD}: 
%Formalism  and twist three distributions,''
Nucl.\ Phys.\ B {\bf 529} (1998) 323.
%%CITATION = HEP-PH 9802299;%%

\bibitem{BB98}
P.\ Ball and V.\ M.\ Braun,
%``Higher twist distribution amplitudes of vector mesons in {QCD}: 
%Twist-4  distributions and meson mass corrections,''
Nucl.\ Phys.\ B {\bf 543} (1999) 201.
%%CITATION = HEP-PH 9810475;%%


\bibitem{ZZC}
A.\ R.\ Zhitnitsky, I.\ R.\ Zhitnitsky and V.\ L.\ Chernyak,
%``QCD Sum Rules And Properties Of Wave Functions Of Nonleading Twist.''
Sov.\ J.\ Nucl.\ Phys.\  {\bf 41} (1985) 284
[Yad.\ Fiz.\  {\bf 41} (1985) 445].
%%CITATION = SJNCA,41,284;%%

\bibitem{BF90}
V.\ M.\ Braun and I.\ E.\ Filyanov,
%``Conformal Invariance And Pion Wave Functions Of Nonleading Twist,''
Z.\ Phys.\ C {\bf 48} (1990) 239
[Sov.\ J.\ Nucl.\ Phys.\  {\bf 52} (1990) 126].
%%CITATION = ZEPYA,C48,239;%%


\bibitem{FGPG72}
S.\ Ferrara {\it et al.},
%``Canonical Scaling And Conformal Invariance,''
Phys.\ Lett.\ B {\bf 38} (1972) 333.
%%CITATION = PHLTA,B38,333;%%

\bibitem{BDKM}
V.\ M.\ Braun {\it et al.},
%``Baryon distribution amplitudes in {QCD},''
Nucl.\ Phys.\ B {\bf 553} (1999) 355.
%%CITATION = HEP-PH 9902375;%%

\bibitem{BKM01}
V.\ M.\ Braun, G.\ P.\ Korchemsky and A.\ N.\ Manashov,
%``Evolution equation for the structure function g2(x,Q**2),''
Nucl.\ Phys.\ B {\bf 603} (2001) 69.
%%CITATION = HEP-PH 0102313;%%


\end{thebibliography}
\end{document}